\documentclass[onecolumn,showpacs,preprintnumbers,amsmath,amssymb,prf]{revtex4}

\usepackage{graphicx}
\graphicspath{{figures/}}
\usepackage{dcolumn}
\usepackage{bm}
\usepackage{xcolor,soul}
\usepackage{subfigure}
\usepackage{floatrow}
\sethlcolor{yellow}

\begin{document}

\preprint{APS/123-QED}

\title{Directional motion of vibrated sessile drops : a quantitative study} 

\author{M. Costalonga$^{1,2}$ and P. Brunet$^{1}$}
\email{philippe.brunet@univ-paris-diderot.fr}
\affiliation{$^{1}$ Laboratoire Mati\`ere et Syst\`emes Complexes UMR CNRS 7057, 10 rue Alice Domon et L\'eonie Duquet 75205 Paris Cedex 13, France.}
\affiliation{$^{2}$ Present address : Department of Mechanical Engineering, Massachusetts Institute of Technology,  77 Massachusetts Avenue, Cambridge MA 02139-4307, USA}

\date{\today}

\begin{abstract}

The directional motion of sessile drops can be induced by slanted mechanical vibrations of the substrate. As previously evidenced \cite{Brunet07,Brunet09,Noblin09}, the mechanical vibrations induce drop deformations which combine axisymmetric and antisymmetric modes.
In this paper, we establish quantitative trends from experiments conducted within a large range of parameters, namely the amplitude $A$ and frequency $f$ of the forcing, the liquid viscosity $\eta$ and the angle between the substrate and the forcing axis $\alpha$. These experiments are carried out on weak-pinning substrates. For most parameters sets, the averaged velocity $<v>$ grows linearly with $A$. We extract the mobility, defined as $s=\frac{\Delta <v>}{\Delta A}$. It is found that $s$ can show a sharp maximal value close to the resonance frequency of the first axisymmetric mode $f_p$. The value of $s$ tends to be almost independent on $\eta$ below 50 cSt, while $s$ decreases significantly for higher $\eta$. Also, it is found that for peculiar sets of parameters, particularly with $f$ far enough from $f_p$, the drop moves in the reverse direction. Finally, we draw a relationship between $<v>$ and the averaged values of the dynamical contact angles at both sides of the drop over one period of oscillation.
\end{abstract}


\maketitle                              

\section{INTRODUCTION}

\begin{figure}
\includegraphics[scale=0.5]{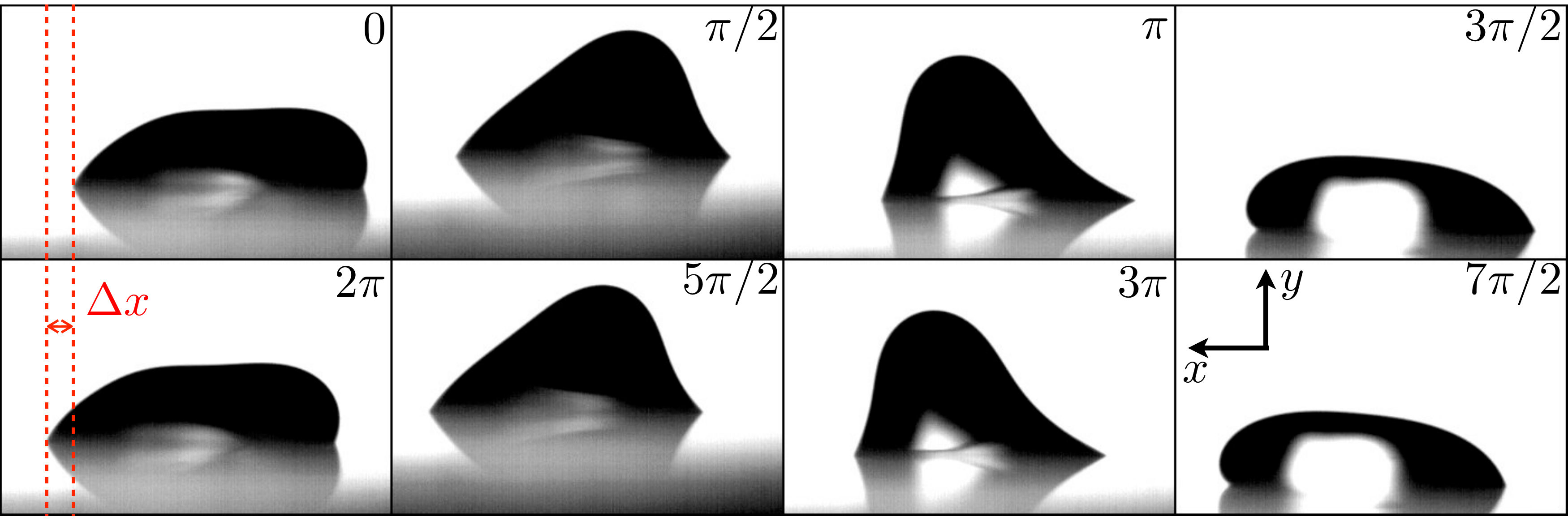}
\caption{Successive deformations of a droplet $V =10 \mu$l of a glycerin/water solution of $\eta$= 31.5 cSt, subjected to a substrate vibration of $f$ = 50 Hz and $A$ = 0.67 mm along a direction of $\alpha$= 60.6$^{\circ}$ with respect to the horizontal. The phase $\phi = \frac{\pi}{2}$ corresponds to the highest vertical and most leftward position of the substrate, while $\phi$ = 0 and $\pi$ correspond to the central position and maximal platform velocity. The drop experiences a net displacement of $\Delta x$ along the $x$ axis, over one period.}
\label{fig:extract_drop}
\end{figure}

The ability to put liquid in motion on a substrate is a challenge in many practical applications, like surface cleaning or homogeneous liquid dispersal. Such handling of small liquid samples is crucial for the development of lab-on-chip platforms for chemical reactions or biological analysis \cite{Blossey}. However, droplet mobility is hindered by pinning forces at the drop contact-line, which originates from physical and chemical imperfections of the substrate, this issue being especially dramatic when the volume of the drop is small enough so that this pinning overcomes most motile forces \cite{Dussan,degennes85}. This microscopic-originating retention is in practice often quantified by the macroscopic apparent receding and advancing angles $\theta_r$ and $\theta_a$, which delimit the range for static equilibrium. In the particular case of drops of colloidal suspensions, pinning at contact-line leads to the (often undesired) formation of 'coffee-stains' during evaporation \cite{Robert1}. These limitations can be tackled with various means, like the use of superhydrophobic surfaces \cite{Marin12,Brunet12}, electrowetting \cite{Mugele08}, a combination of both \cite{Lapierre} or Leidenfrost drops on ratcheted surfaces \cite{Linke06,Lagubeau11}. 

Several studies have evidenced that mechanical vibrations of the substrate can induce drop motion \cite{Daniel05,Dongetal07,Brunet07,Brunet09,Noblin09,Mettu_Chaudhury11,Borcia14,Sartori15,Sartori19}, as well as particle resuspension \cite{Whitehill10}. These phenomena have generic features, in common with other related situations : the generation of surface waves on thin films by external vibrations \cite{Oron13}, the non-harmonic response of drop subjected to MHz ultrasonic surface waves inducing both their motion and low-frequency oscillations \cite{Brunet10}, or the crawling motion of sessile droplets on solid surfaces via acoustic radiation pressure \cite{Alzuaga05}. 

Figure \ref{fig:extract_drop} shows the shapes and positions of a drop on a horizontal substrate subjected to slanted unbiased vibrations. Pictures are taken at successive phases from 0 to $7 \pi /2$. As the drop responds harmonically, the successive shapes of its free-surface are identical with each other at the same phase (mod. $[2 \pi]$) of the forcing. A net motion $\Delta x$  appears over one period, from right to left on Figure \ref{fig:extract_drop}. During one period, the drop free-surface experiences strong and asymmetric deformations, which induce unbalanced Young forces at contact-lines \cite{Brunet07}. The left and right positions of the contact-line show back-and-forth motion, often of much faster velocity than the averaged velocity of the rectified motion of the droplet.  

The aim of the present study is to seek for constitutive laws to enable a better understanding of the underlying mechanisms. To do so, we quantify the influence of the main control parameters and of physical quantities on the droplet response to vibrations and its resulting net motion.  We present extensive results within a large span of frequency, amplitude of vibrations, angle between the substrate and axis of vibration, drop volume and liquid viscosity. We extract the amplitude threshold for drop motion, and the drop mobility quantified by the relationship between the net averaged velocity $< v >$ and the amplitude or acceleration of the forcing. Importantly, we carry out experiments with low friction substrate (small contact angle hysteresis (CAH)), which lowers the acceleration threshold to observe droplet's motion, together with facilitating quantitative comparison with existing models.

One of the surprising features of the directional motion is that it does not require bias in the vibration nor anisotropy in surface texture or chemistry. Previous studies \cite{Brunet07,Brunet09} showed that if the vibration is slanted with respect to the substrate, the drop can respond with both symmetrical and asymmetrical modes, and underlined the importance of the coupling between both modes. This was further confirmed by Noblin \textit{et al.} \cite{Noblin09}, who operated with two decoupled vibrating benches and emphasized the role of phase shift between the symmetric and asymmetric forcing. Very recently, Sartori \textit{et al.} \cite{Sartori19} showed a strong correlation between the direction and velocity of motion, and the phase-shift between the forcing vibration and the oscillations of the basal radius.

Therefore, both the asymmetric "rocking" and symmetric "pumping" modes are required to produce a directional motion. Indeed, while the rocking mode allows for the dynamical contact angle to reach values beyond the range of CAH $[\theta_r , \theta_a ]$, the symmetric mode prescribes a modulation of the height and base radius, as shown on Figure \ref{fig:extract_drop}, where the drop is successively flattened and pushed rightward, then stretched and pushed leftward.  It turns out that a stretched drop is more compliant to lateral forcing than a flattened one \cite{Brunet07}, which enhances the asymmetry and enables a nonzero averaged lateral force on the drop. The resulting motion over one period is to the left. If submitted to the same forcing, a pendant drop down the substrate would move to the other direction \cite{Brunet07,Brunet09}.

In order to explain and quantitatively predict this directional motion, theoretical and numerical studies were proposed, mostly in situations of vertically vibrated drops climbing up inclines : Benilov and co-authors first considered the case of a two-dimensional (2D) flat viscous drop \cite{Benilov10} responding quasi-statically to periodic forcing, then that of flat inviscid drops \cite{Benilov11} and finally of thick drops (static contact-angle $\theta_S  \gg $0) \cite{Benilov13}. The main conclusions of the studies were that : (1) some inertia is required to obtain realistic (i.e. comparable to experiments) acceleration threshold for climbing \cite{Benilov11} and (2) $\theta_S$ plays a crucial role in the climbing threshold, generally a large $\theta_S$ favours climbing for low enough frequency whereas it penalises climbing for high enough frequency. At odds with the aforementioned approaches, John and Thiele \cite{John_Thiele10} addressed the problem of a flat climbing drop using lubrication approximation, especially in the quasi-static limit (i.e. low frequency: the drop responds in phase with the vibrations). Their model could capture realistic orders of magnitude for climbing. In a very recent paper, Bradshaw and Billingham \cite{Bradshaw_Billingham18} investigated the situation of thick inviscid drops, where the effect of viscosity was embedded in the relationship between the dynamical contact-angle and the contact-line velocity, also including a pinning force due to CAH. Following a previous study \cite{Bradshaw_Billingham16} in the situation of thin tridimensional (3D) drops, the far-reaching results in \cite{Bradshaw_Billingham18} showed realistic trends, especially concerning the non-trivial influence of the forcing frequency and the CAH. While all of these studies were carried out with bidimensional (2D) drops, Ding \textit{et al.}'s study  \cite{Ding18} considered 3D drops with full Navier-Stokes equations and diffuse interface model to account for free-surface and contact-line dynamics. They obtained shapes with unprecedented realism and quantitative trends that suggested the importance of the nonlinear response of the wetted area over one period. 

Also recently, Borcia \textit{et al.} \cite{Borcia14}, Sartori \textit{et al.} \cite{Sartori15} addressed the vibration-induced motion of two-dimensional sessile drops with phase-field numerical methods, with comparisons to experiments. Although CAH could not be included within these models, they could capture realistic behavior, especially in the influence of viscosity, wetting-conditions \cite{Borcia14}, or even the occurrence of parametric forcing with a drop responding at half the forcing frequency \cite{Sartori15}.

Based on the compared analyses of the aforementioned (sometimes contradictory) approaches, the need for exhaustive and quantitative experimental results to validate the aforementioned models, is crucial. Especially, two main issues remain and our experiments aim to provide insights to address them :

- First, the relative importance of viscosity and inertia is unclear. The aforementioned theoretical studies propose two opposed models where either viscosity is neglected \cite{Benilov10,Benilov11,Benilov13} or embedded in a contact-line friction \cite{Bradshaw_Billingham18,Bradshaw_Billingham16}, or inertia is neglected \cite{John_Thiele10}. While the sequence of figure \ref{fig:extract_drop} suggests that a phase shift between the substrate vibrations and the drop deformations is required to produce an averaged asymmetry, it has been shown that very viscous drops oscillating in phase with the substrate also exhibit a net mean motion \cite{John_Thiele10}, although at much smaller velocity. 

- Second, it is unclear to what extent one can relate the averaged unbalanced Young's force - based on the contact-angles at the front and the rear of the drop, to the net velocity. Previous measurements showed an empirical relationship between the capillary number Ca and the unbalanced Young force evaluated over one period \cite{Brunet07}, but this point required further confirmations over a larger range of parameters and with conditions closer to an ideal situation, in which the CAH-based friction force could be as small as possible \cite{Sartori19}.

The paper is organized as follows : Section II describes the experimental setup. Section III illustrates qualitatively the phenomenon under study. Section IV presents the different experimental results. Section V proposes some discussion and interpretations of the results and conclude about the main trends.

\section{EXPERIMENTAL SETUP}

\begin{figure}
\includegraphics[scale=0.5]{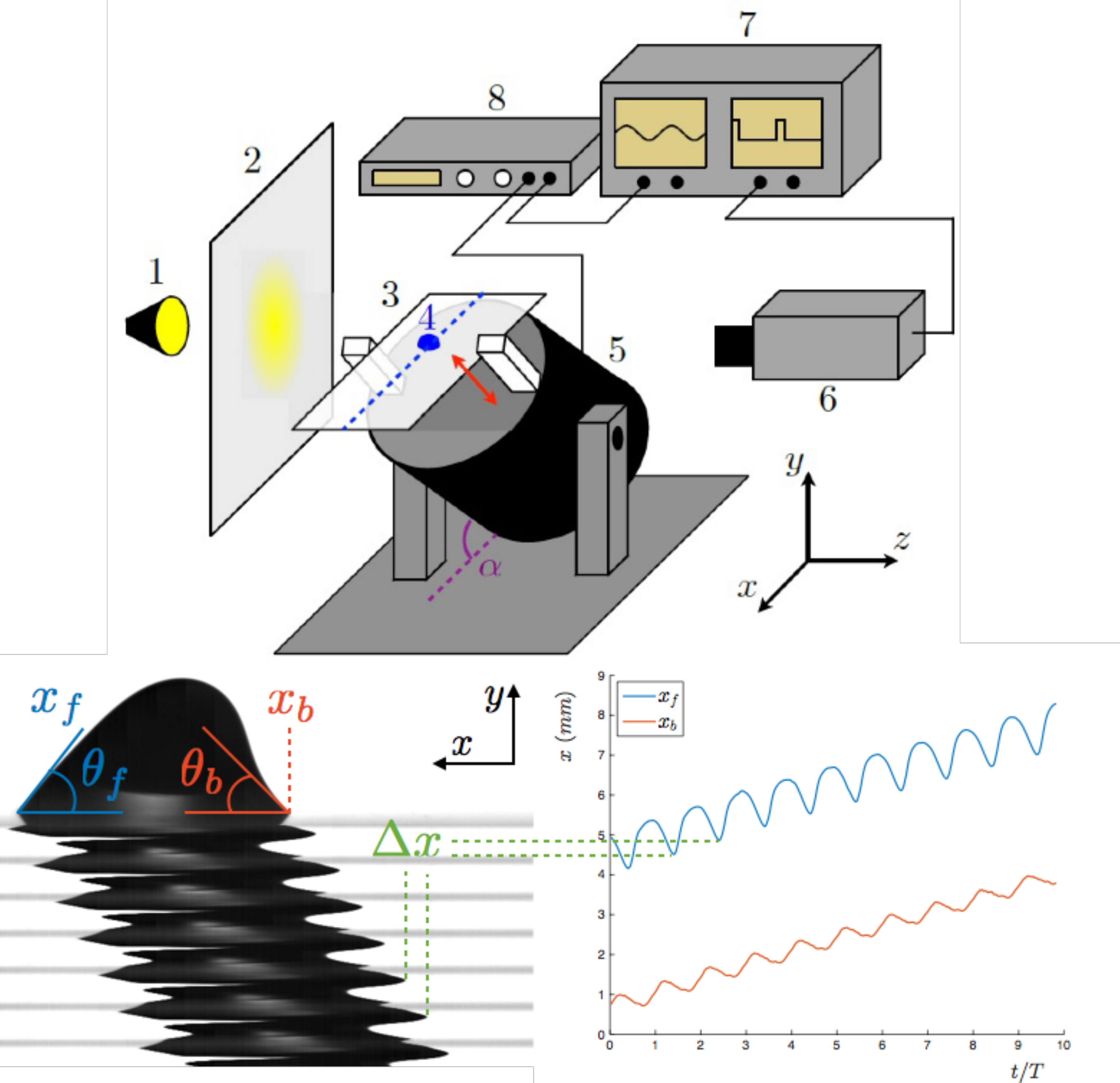}
\caption{\textit{Top} - Sketch of the experimental setup: a droplet of glycerin/water mixture (4) is deposited on a chemically-treated glass slide (3), which is fixed on an inclinable vibration generator (5). Vibrations are prescribed by a function generator (7) with an amplifier (8), that delivers a signal synchronized to a high-speed camera (6) with optical zooms of various magnification. The drop is lighted from the back by a halogen lamp (1) projected on a diffuser screen (2). \textit{Bottom left} - Definition of time-varying drop contact-line positions $x_f$ and $x_b$ and dynamical contact angles $\theta_f$ and $\theta_b$. In this example, the drop motion is from right to left. \textit{Bottom right} - An example of contact-line positions $x_f$ and $x_b$ as a function of time, showing the superimposition of a periodic back-and-forth motion and a directional averaged one.}
\label{fig:setup_vibrations2}
\end{figure}

The experimental setup is depicted in Figure \ref{fig:setup_vibrations2}-(\textit{Top}). A function generator (7) (Rigol - A4162) together with a power amplifier (8) (Labworks PA141) prescribe a time-periodic signal to a vibration generator (5) (Labworks ET 141). The mechanical vibration is transmitted to an axis on which a substrate (glass slide (3)) is mounted and glued with epoxy resist on a plexiglass plate attached to the vibration axis. The amplitude and frequency of vibrations are the main control parameters of the experiment. The plate displacement over time is then $A cos (\omega t)$ along the axis of vibration. This leads to a periodic acceleration through the same direction in phase opposition with the displacement, which maximal value $a = A \omega^2$. The axis of vibration makes an angle $\alpha$ with the horizontal, that is varied within $[0^{\circ}, 90^{\circ}]$. 

A droplet of volume $V$ (4) is gently deposited on the substrate. The liquid is a mixture of water and glycerin, with percentage in mass of glycerin varying from 40 to 90 \%, resulting in a kinematic viscosity $\eta$ varying from 5.9 to 191 cSt, but weakly varying density $\rho$ and surface tension $\gamma$. Within this range, the viscosity is large enough to prevent the splitting of the drop under vibrations. We dismissed pure glycerin which shows significant drift (decrease) in viscosity when exposed to ambient air during the time of experiments, as it quickly absorbs ambient water vapor.

The volume $V$ is chosen small enough for the drop to adopt the shape of a spherical cap, at rest : the static Bond number Bo measuring the relative magnitude of gravity and capillarity forces, $\text{Bo} = \rho g V^{2/3}/\gamma$ remains smaller than one. In practice, we opted for a volume of 10 $\mu$l (Bo between 0.65 and 0.84, depending on the percentage of glycerin) for most of the experiments reported in this paper. It will be explicitly mentioned when experiments are carried out at different volumes.

The droplet motion and deformations are recorded either from the side or from above, with a high-speed camera (Photron SA3 (6)) operating at 1000 or 2000 fps and with optical zooms together with extension tubes, both ensuring a magnification of a few microns per pixel (for the recording of local deformations) to a few tens of microns per pixel (for the recording of the drop net motion). Figure \ref{fig:setup_vibrations2}-(\textit{Bottom, Left}) shows the side view of the drop, and its successive footprints taken after several periods, and at the same phase, with the absolute positions of the left (front) and right (back) contact-lines ($x_f$) and ($x_b$), the corresponding dynamical contact angles (CA) ($\theta_f$) and ($\theta_b$) and the net displacement during each period $\Delta x$. Figure \ref{fig:setup_vibrations2}-(\textit{Bottom, Right}) shows an example of a time-evolution of ($x_f$) and ($x_b$), evidencing both the back-and-forth motion and the net displacement with averaged velocity $<v> = \Delta x . f$.

The substrate are glass slides coated with a Self-Assembled Monolayer (SAM) of a low surface energy fluoropolymer. Octadecyl-trichlorosiloxane (OTS) makes covalent bonds on activated Si-O. The activation is ensured by Oxygen Plasma treatment, and the glass slides are then kept several hours in a solution 10$^{-3}$ M of OTS in Hexane. The whole process is carried out in a class 1000 clean room. This treatment allows for weak retention force with reproducible, homogeneous and long-lasting properties. The resulting advancing and receding CAs are respectively: $\theta_a$ = 107$^{\circ}$ and $\theta_r$ = 86$^{\circ}$. 

\section{DROPLET DYNAMICS : A QUALITATIVE DESCRIPTION}

Since directional displacement originates from a time-averaged symmetry breaking in the drop shape, it is of primary importance to characterize how the free-surface responds to external vibrations. From the pioneering studies of Rayleigh and Lamb \cite{Lamb}, inertio-capillary modes of a freely suspended sphere have eigenfrequency $f_n$ :

\begin{equation}   
f_n =  \frac{1}{2 \pi} \left( \frac{n (n-1) (n+2) \gamma}{\rho R^3} \right)^{\frac{1}{2}}
\label{eq:ray}
\end{equation}

\noindent with $n$ is the mode number and $R$ is the drop radius.  

However, real situations involve more complex effects, in which several questions remain unanswered \cite{Dongetal07,Tsamopoulos_Brown83,Strani_Sabetta84,Celestini06,Bostwick_Steen09, 
Fayzrakhmanova_Straube09,Smith10,Brunet_Snoeijer11,Savva_Kalliadasis_JFM13,Savva_Kalliadasis_JFM14,Chang_Bostwick_Steen13,Bostwick_Steen_JFM14,Bostwick_Steen_JFM15,Sharp12}. Especially, our situation of a sessile drop shows qualitative and quantitative differences with eq. (\ref{eq:ray}) \cite{Dongetal07,Strani_Sabetta84,Celestini06,Fayzrakhmanova_Straube09,Smith10,Savva_Kalliadasis_JFM13,Savva_Kalliadasis_JFM14,Chang_Bostwick_Steen13,Bostwick_Steen_JFM14,Bostwick_Steen_JFM15}. First, the contact with the substrate enables a non-degenerated translational mode (n=1) with finite frequency \cite{Strani_Sabetta84,Dongetal07}. This is the \textit{rocking mode} (i.e. the drop rocks from left to right), excited when the drop is subjected to lateral forces. The eigenfrequency of the rocking mode is also proportional to $\left(\frac{\gamma}{\rho V}\right)^{\frac{1}{2}}$ \cite{Dongetal07} with a prefactor depending on the wetting conditions \cite{Celestini06,Sharp12}. Secondly, the complex dynamics of contact-lines, involving significant pinning force on real substrates, can generate stop-and-go dynamics \cite{Noblin04}. Generally, the problem is treated by prescribing ad-hoc conditions, that relate the instantaneous contact-line velocity and the macroscopic deformation at the vicinity of the contact-line, as stated for instance in \cite{Bradshaw_Billingham18,Bradshaw_Billingham16,Perlin95,Fayzrakhmanova_Straube09}. Our choice of a low-friction substrate aims to minimise this complexity as much as possible. Thirdly, the finite value of the frequency prescribes that within a (unsteady) viscous BL of thickness $\delta = \left(\frac{2 \nu}{\omega}\right)^{1/2}$, the motion of fluid is in phase with that of the substrate. Above this layer, the fluid follows the plate oscillations, but with a phase lag due to inertia, which can influence in turn the motion of contact-lines $x_f(t)$ and $x_b(t)$. Therefore, substrate-induced constraints influence the symmetric Rayleigh-Lamb modes and the values of $f_n$ \cite{Strani_Sabetta84,Bostwick_Steen09,Sharp12}. 

\begin{figure}
\includegraphics[scale=0.5]{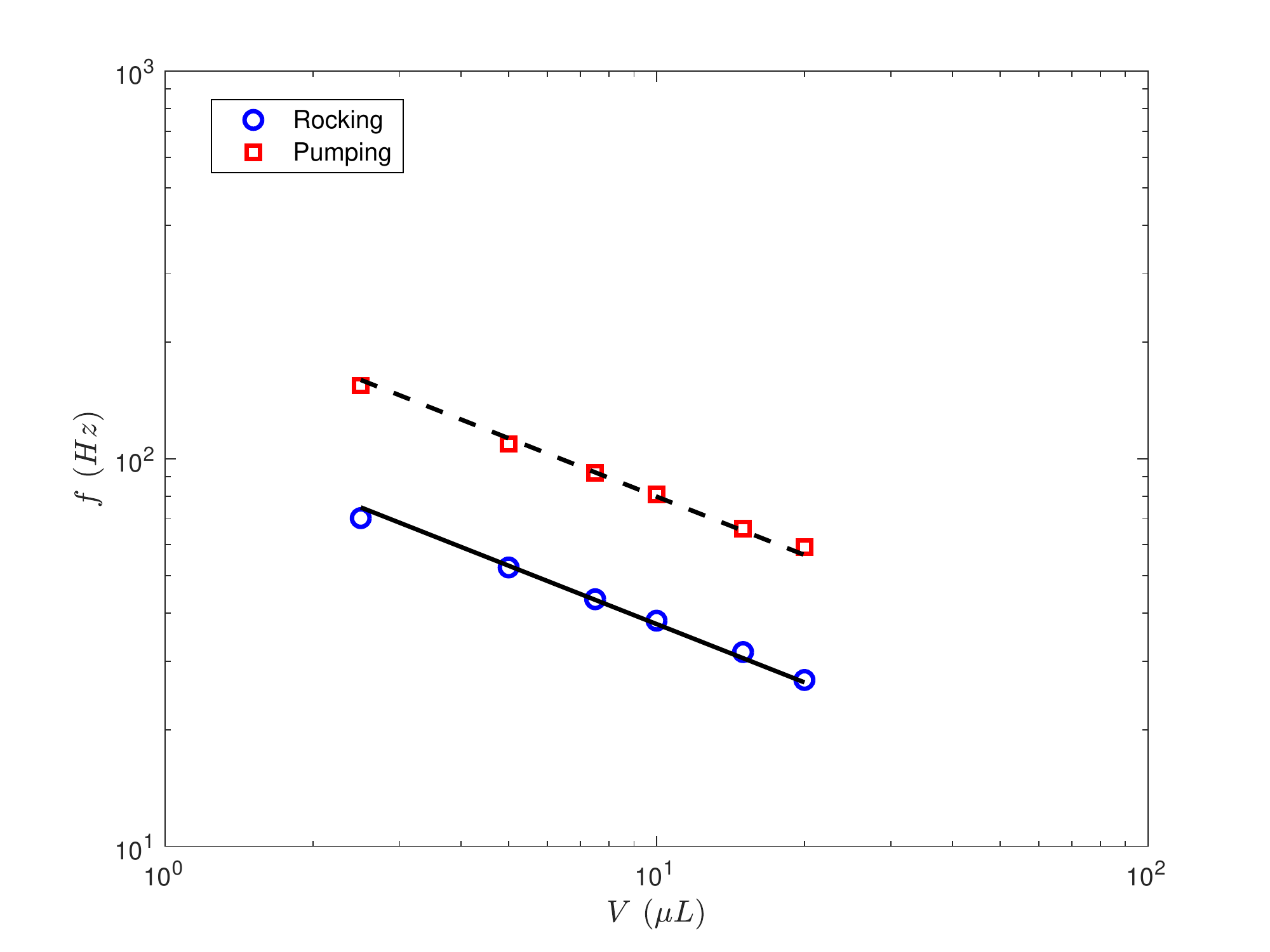}
\caption{Eigenfrequencies of symmetric pumping mode ($n$=2) and asymmetric rocking mode ($n$=1) versus drop volume, deduced from transient relaxation dynamics of drops after an initial kick. The lines represent a fit by the power law $f \sim V^{-1/2}$.}
\label{fig:f_eigenmodes}
\end{figure}

We measure the frequencies of the two first eigenmodes by recording the transient response of the free-surface deformations following a "kick", i.e. a step of acceleration by the shaker. To excite independently the pumping and rocking modes, we impose either vertical ($\alpha$ = 90$^{\circ}$) or horizontal ($\alpha$ = 0$^{\circ}$) forcing. We take drop volume $V$ in a range encompassing widely the values taken in experiments. Figure \ref{fig:f_eigenmodes} shows the eigenfrequencies versus $V$ for the lowest order pumping and rocking modes, which confirms the decrease of $f$ with $V$ via a power-law of exponent $-1/2$. The prefactor is smaller for the rocking mode (asymmetrical, $n$=1) than for the pumping mode (symmetrical, $n$=2).

The eigen frequency for the first asymmetric rocking mode is then empirically determined by:

\begin{equation}
f_r = \beta_r . \left( \frac{\gamma}{\rho V} \right)^{\frac{1}{2}}
\label{eq:fr}
\end{equation}

\noindent and for the first symmetric pumping mode, it yields a similar relationship:

\begin{equation}
f_p = \beta_p . \left( \frac{\gamma}{\rho V} \right)^{\frac{1}{2}}
\label{eq:fp}
\end{equation}

The coefficients $\beta_r$ and $\beta_p$ are respectively equal to 0.95$\pm$0.01 and 0.47$\pm$0.01. As stated above, they should depend on the wetting conditions \cite{Celestini06,Sharp12}, an effect which is is not addressed in this study and to some extent on the viscosity $\eta$ \cite{Sharp12}.

The sequence in Fig.~\ref{fig:extract_drop} represents typical drop deformations under moderate forcing. Clearly, the drop shape responds with a phase lag with respect to the excitation. The drop shows maximal vertical stretching between $\pi /2$ and $\pi$, while the acceleration is maximal in the downward and leftward direction for $\pi /2$. \textit{A priori}, the phase lag can be different for the pumping and the rocking modes : the drop presents the most left/right-asymmetric shape at a phase slightly after $\pi$, i.e. after the phase related to the maximal upward stretching. It is related to that the thickness of the unsteady BL $\delta$ is here smaller than the drop height $h$. Hence, to quantify the importance of inertia, one can build a dimensionless BL thickness :

\begin{equation}
\delta^* = \frac{\delta}{h} = \left( \frac{\nu}{\pi f h^2} \right)^{\frac{1}{2}}
\label{eq:boundarylayer}
\end{equation}

If $\delta^* \gg$1, the shear from the substrate vibration is entirely diffused in the liquid during a period of oscillation : the whole drop responds in phase with the forcing. Conversely if $\delta^* \ll$1, only a thin layer near the liquid/solid interface, moves in phase with the plate. As our study aims to quantify the influence of viscosity, our experimental range shall include these two extreme situations, and of course the range in between, where $\delta^*$ is of the order of one. 

Another dimensionless number which is susceptible to influence the phase shift between excitation and drop response, is the ratio between the excitation frequency and the frequency of the first symmetric eigenmode, namely $f_p$ :

\begin{equation}
f^* = \frac{f}{f_p}
\label{eq:fstar}
\end{equation}

The Ohnesorge number quantifies the relative importance of viscous and capillary effects in free-surface dynamics :

\begin{equation}
\text{Oh} = \nu \left(\frac{\rho}{\gamma R} \right)^{\frac{1}{2}}
\label{eq:Oh}
\end{equation}

\noindent which for drop of typical volume 10 $\mu$l ($R \simeq$ 1.68 mm) and the water/glycerin mixtures of various composition, ranges from 0.0146 to 0.66.

\section{QUANTITATIVE RESULTS}

Previous experiments \cite{Brunet07,Brunet09,Noblin09,Sartori15,Sartori19} and numerical simulations \cite{Bradshaw_Billingham16,Bradshaw_Billingham18,Ding18} showed that directional motion is induced providing the amplitude $A$ (or maximal acceleration $a$) gets stronger than a threshold $A_{th}$ (resp. $a_{th}$, and that the averaged velocity $<v>$ generally increases with the forcing. The threshold originates from the CAH due to substrate imperfections. Overall, one can set a general empirical law for the averaged velocity :

\begin{equation}
<v> = s \times (A - A_{\text{th}})^{\chi}
\label{eq:vmoy}
\end{equation}

\noindent where $s$ can depend on viscosity, frequency, inclination angle $\alpha$ and wetting conditions. The exponent $\chi$ was found to be roughly equal to one in previous experiments \cite{Brunet07,Noblin09,Sartori15,Sartori19}, but more recent numerical results showed better agreement with a quadratic behavior ($\chi$ = 2) for eq.~(\ref{eq:vmoy}), in the situation of weak forcing and low CAH \textcolor{blue}{\cite{Benilov10,Benilov11,Benilov13,Bradshaw_Billingham16,Bradshaw_Billingham18}}, and a crossover toward a linear behavior ($\chi$ = 1) at stronger forcing and/or larger CAH \cite{Ding18}. Other numerical results with no CAH were consistent with $\chi$ between 1.5 and 2, the exponent being dependent on the CA \cite{Borcia14}. On very slippery surfaces, it was even found a saturation and decrease of $<v>$ with $A$ at strong forcing \cite{Sartori19}.

Given the relative discrepancy between the different previous results, our experiments aim to extract the values of $s$, $A_{\text{th}}$ and $\chi$ with different values of the aforementioned parameters. 

\subsection{The influence of the angle between substrate and vibrations}

\begin{figure}
(a)\includegraphics[scale=0.55]{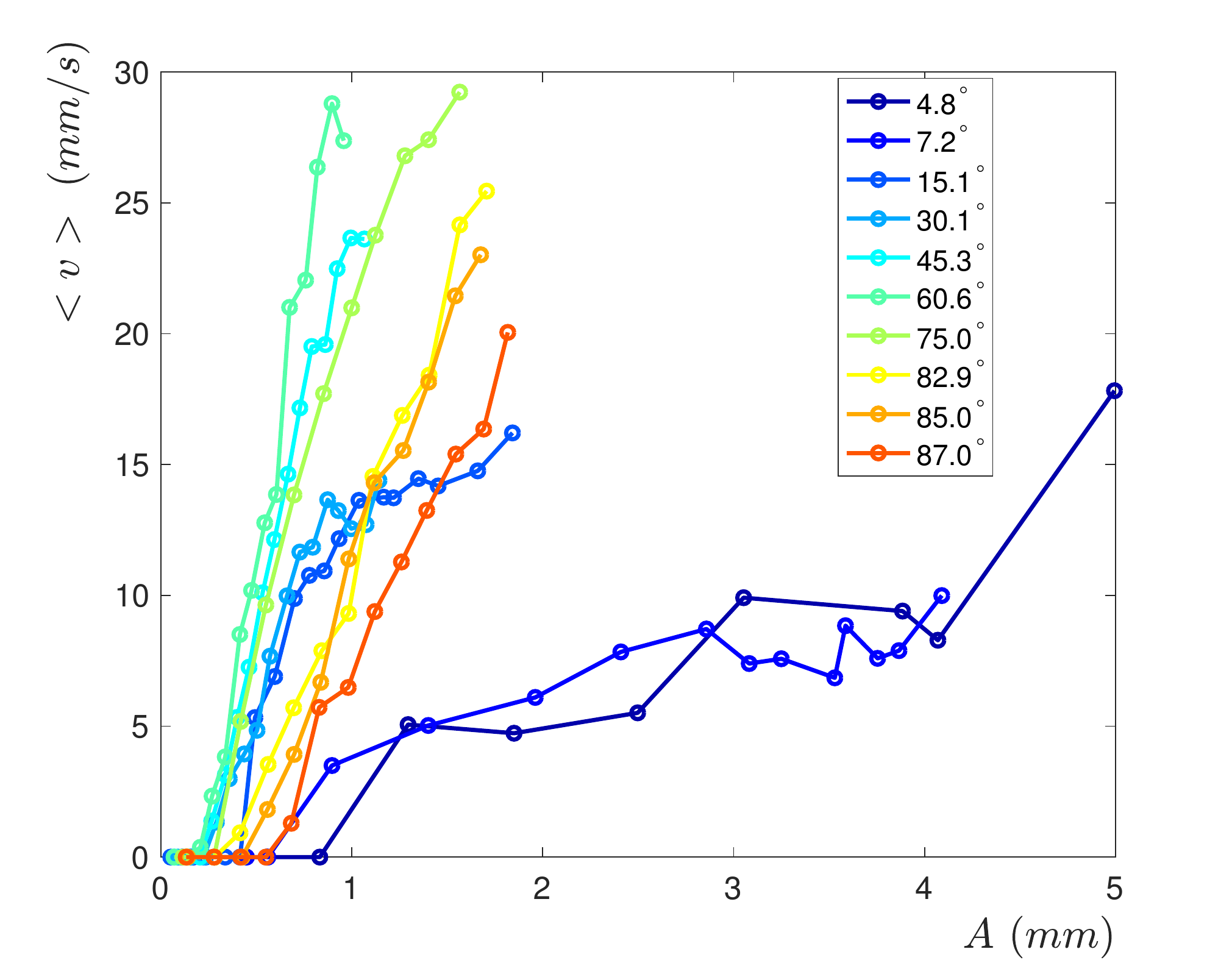} \\
(b)\includegraphics[scale=0.32]{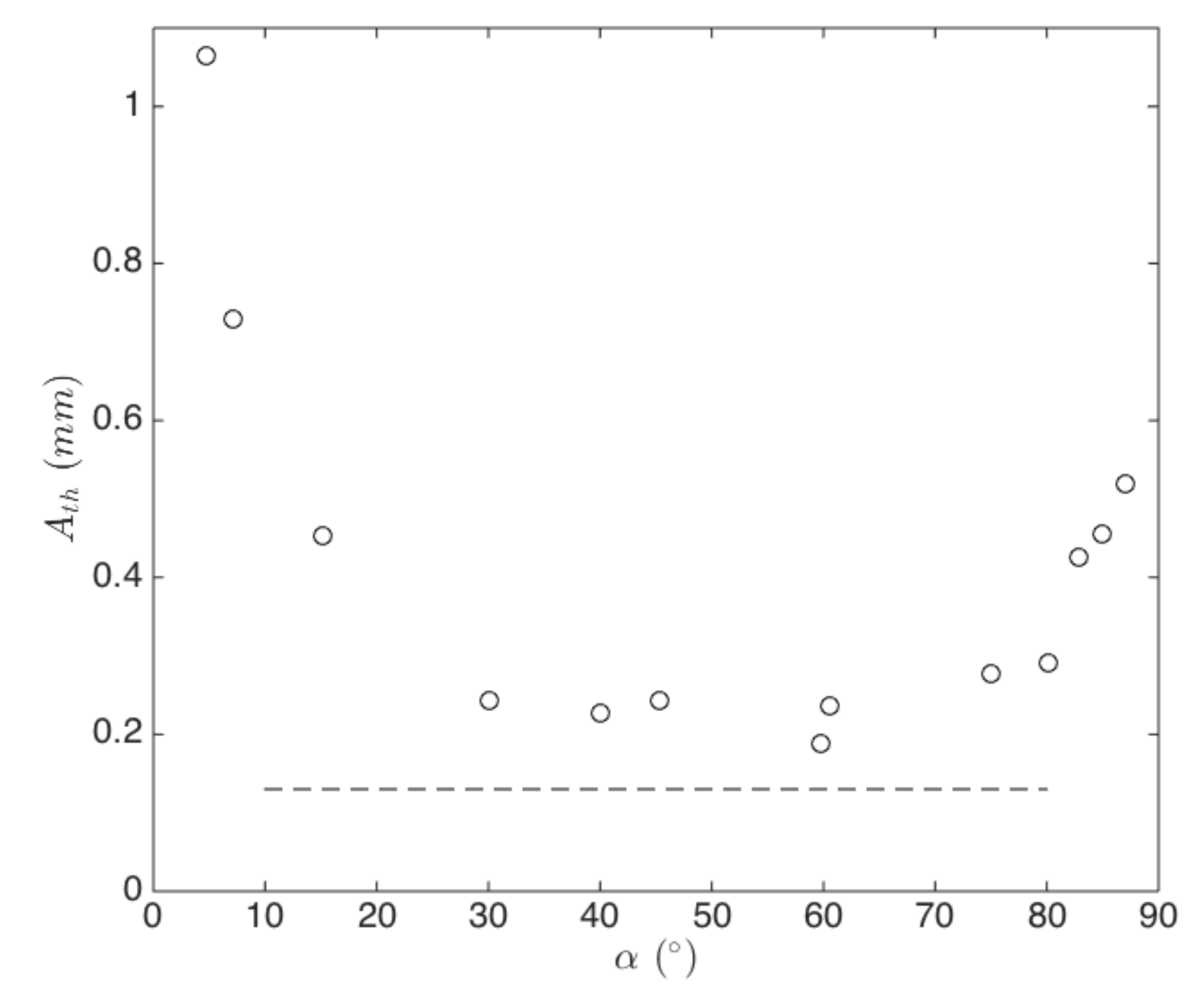}
(c)\includegraphics[scale=0.45]{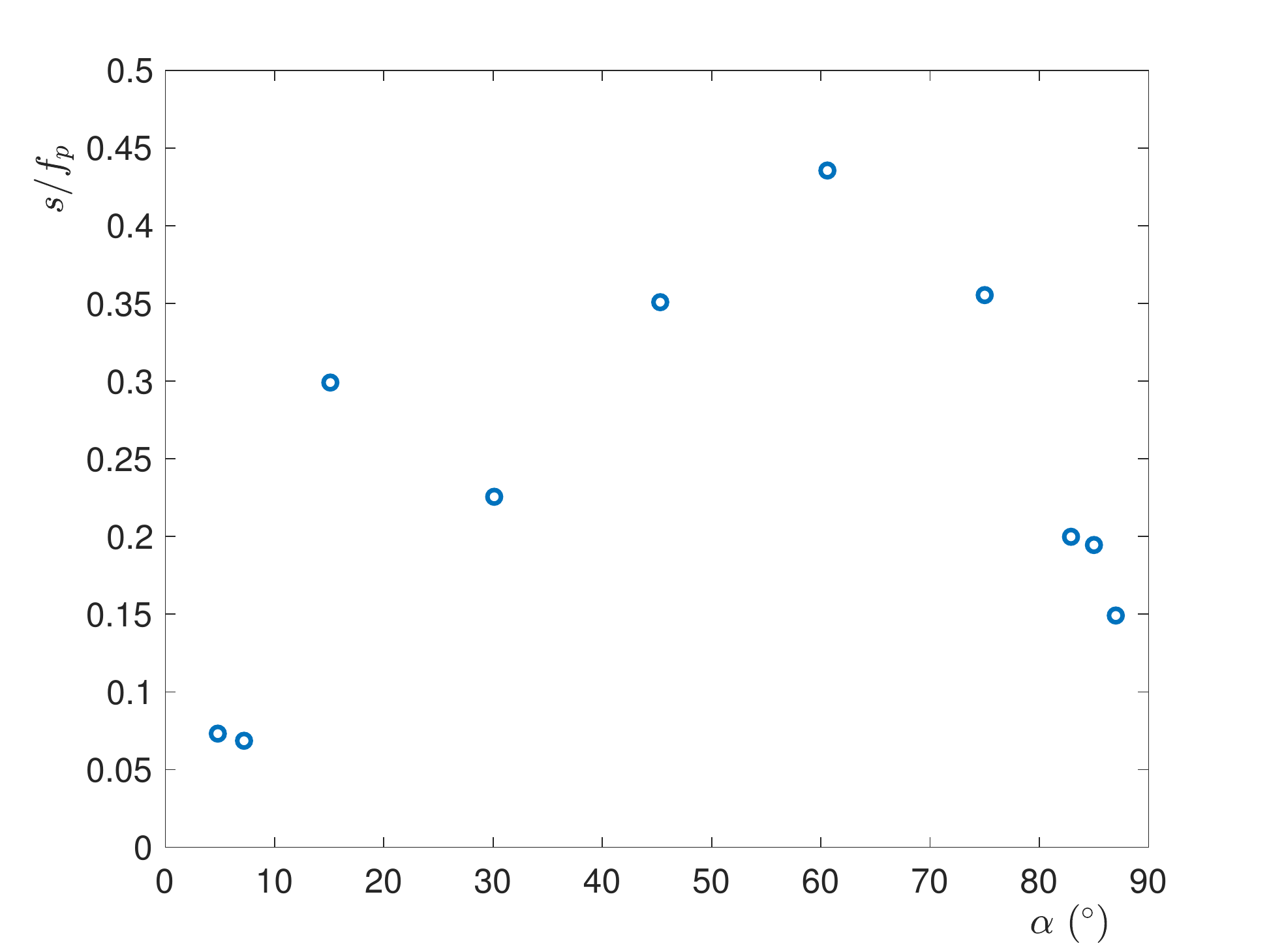}
\caption{(a) Time-averaged velocity of the drop center of mass versus amplitude of vibration, for different angles $\alpha$ between the horizontal and the axis of vibrations. $f$ = 50 Hz, $V$ = 10 $\mu$l and $\eta$ = 31.5 cSt. (b) The threshold $a_{\text{th}}$ divided by $g$ and (c) the coefficient $s$ of the linear variation of the velocity with $A$ plotted versus $\alpha$ (same parameters as (a) otherwise).}
\label{fig:influ_inclin}
\end{figure}

We first address the dependence of $<v>$ with respect to the slant angle $\alpha$. Figure \ref{fig:influ_inclin}-(a) shows the mean drop velocity versus amplitude $A$, for a frequency of 50 Hz and for several values for the angle $\alpha$. It appears that data are well fitted by eq.~(\ref{eq:vmoy}), with $\chi$=1. One extracts the values of both $s$ and $A_{\text{th}}$, versus $\alpha$ (Fig.~\ref{fig:influ_inclin}-(b)). In Figure \ref{fig:influ_inclin}-(c), the coefficient $s$ is plotted versus $\alpha$. This suggests that the drop mobility shows an optimum at $\alpha \simeq$ 60$^{\circ}$. Still, even if a sharp decrease of mobility is observed as $\alpha \rightarrow$ 0 and $\alpha \rightarrow \pi /2$, it is still possible to move drops by moderate vertical vibrations even for nearly vertical or horizontal substrates.

Considering that both symmetric and asymmetric modes are required to produce a net motion, it is indeed expected that as $\alpha$ approaches 0 or $\frac{\pi}{2}$, $A_{\text{th}}$ strongly increases, and actually almost diverges, and also that $s$ strongly decreases toward zero. Indeed for $\alpha = \frac{\pi}{2}$, only the axisymmetric pumping mode is excited. Hence, although the contact-line can be unpinned, there is no way for the dynamical contact-angles to be simultaneously larger than $\theta_a$ on the one side of the drop and smaller than $\theta_r$ on the other side. On the other hand for $\alpha =$0, only the rocking mode is excited: although the drop center-of-mass and the left and right positions of the contact-line move back-and-forth at the prescribed frequency, the time-averaged asymmetry of the drop shape is null. No net motion of the drop can be noticed over several periods. Hence, a combination of both modes is required for a directional motion.

Another interesting behavior is the quasi independence of $A_{\text{th}}$ within the range $\alpha \in [20^{\circ};80^{\circ}]$ (see Fig.~\ref{fig:influ_inclin}-(b)). Comparing the threshold in acceleration ($a_{\text{th}}$)  to the effective capillary force build on the CAH, $F_{\text{cap}} = \gamma \pi V^{1/3} (\cos \theta_r - \cos \theta_a)$, and represented by the dashed line in Fig.~\ref{fig:influ_inclin}-(b), it turns out that it is slightly above this characteristic static value, and it is approximately equal to 2.2 times the gravity acceleration. Above 80$^{\circ}$, $a_{\text{th}}$ sharply increases with $\alpha$, and below 20$^{\circ}$ the increase is even sharper. Let us mention that for $\alpha$ below roughly 5$^{\circ}$, the accuracy of measurements is impacted by the limitation of our vibrating bench : the motion departs from a purely rectilinear motion, especially at high amplitude.

No simple explanation can be proposed for the maximum being at 60$^{\circ}$. As stated above, the drop requires both pumping and rocking modes to be excited with large enough amplitudes; therefore the most natural intuition would have suggested this maximum to be around 45$^{\circ}$. Though, this maximum is found to be of a rather plat profile around 60$^{\circ}$ and, similarly to $A_{th}$, the variation of $s$ with $\alpha$ is relatively weak between 20$^{\circ}$ and 80$^{\circ}$.

\subsection{The influence of the excitation frequency}

\subsubsection{Backward motion for specific values of $f^*$}

Vibration-induced directional motion of drops has been observed over a large range of excitation frequency $f$ \cite{Brunet07,Brunet09,Noblin09,Sartori15,Sartori19}, providing $f$ is of the same order of magnitude as the first eigenfrequencies of the inertia-capillary modes of the drop, namely $f_r$ and $f_p$, and potentially $f_{2p}$ and $f_{2r}$ the resonance frequencies of the 2$^{\text{nd}}$ order pumping and rocking modes. 
For this reason, the value of volume $V$ is supposed to rule the dependence of $<v>$ (or $s$ and $A_{\text{th}}$ on $f$), as both $f_r$ and $f_p$ are dependent on $V$, see eqs.~(\ref{eq:fr}) and (\ref{eq:fp}). Hence, we will consider the dimensionless excitation frequency $f^*$ defined in eq.~(\ref{eq:fstar}).

\begin{figure}
\subfigure[]{\includegraphics[scale=0.43]{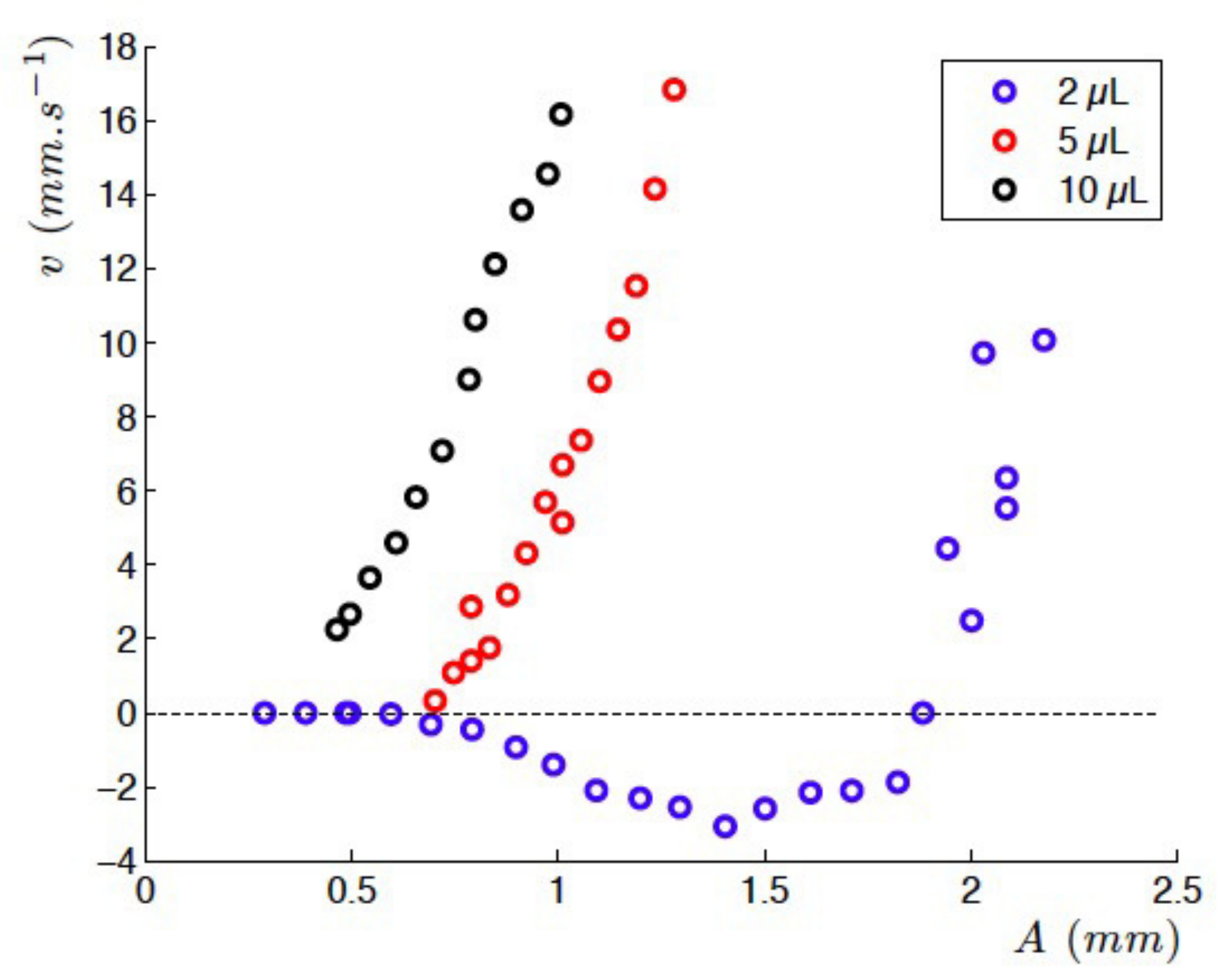}}
\subfigure[]{\includegraphics[scale=0.43]{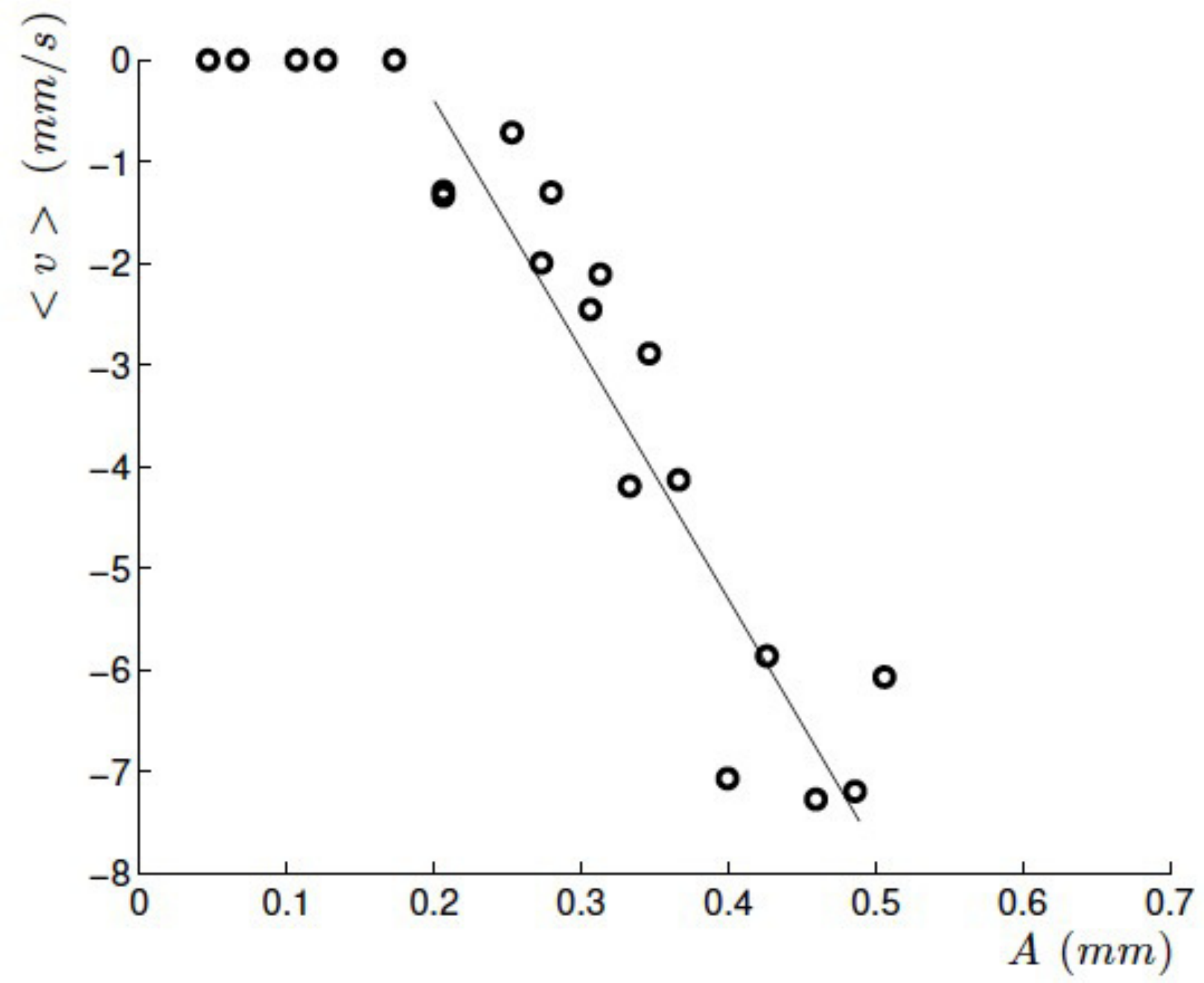}}
\caption{(a) Drop velocity versus amplitude under the same vibration frequency ($f$=40 Hz) and liquid viscosity ($\eta$ = 28.8 cSt), with three different drop volume $V$ = 2, 5 and 10 $\mu$l. The dimensionless frequency $f^*$ is respectively 0.22, 0.37 and 0.49. For $f^*$=0.22, a backward motion is observed within a large range of amplitude. (b) Drop velocity versus amplitude in a situation of backward motion. Drop volume $V$=10 $\mu$l, $f$ = 120 Hz ($f^*$ = 1.48) and $\eta$ = 7.1 cSt.}
\label{fig:influ_freq_intro}
\end{figure}

We first underline the influence of $V$ (through $f^*$) on the drop velocity. Figure \ref{fig:influ_freq_intro}-(a) shows the averaged velocity $< v >$ versus $A$ for three different volumes 2, 5 and 10 $\mu$l. It is striking that $V$ has a stark influence on $< v >$ : for the same forcing $(A,f)$, not only the onset of directional motion changes, but also the direction of the motion can be reversed. The backward motion corresponds to a drop of 2 $\mu$l and $f$ = 40 Hz ($f^*$= 0.22), and is typical for small values of $f^*$ and moderate $A$. It is also observed for different $f$ and $V$ providing $f^* <$ 0.25. 

\begin{figure}
\includegraphics[scale=0.40]{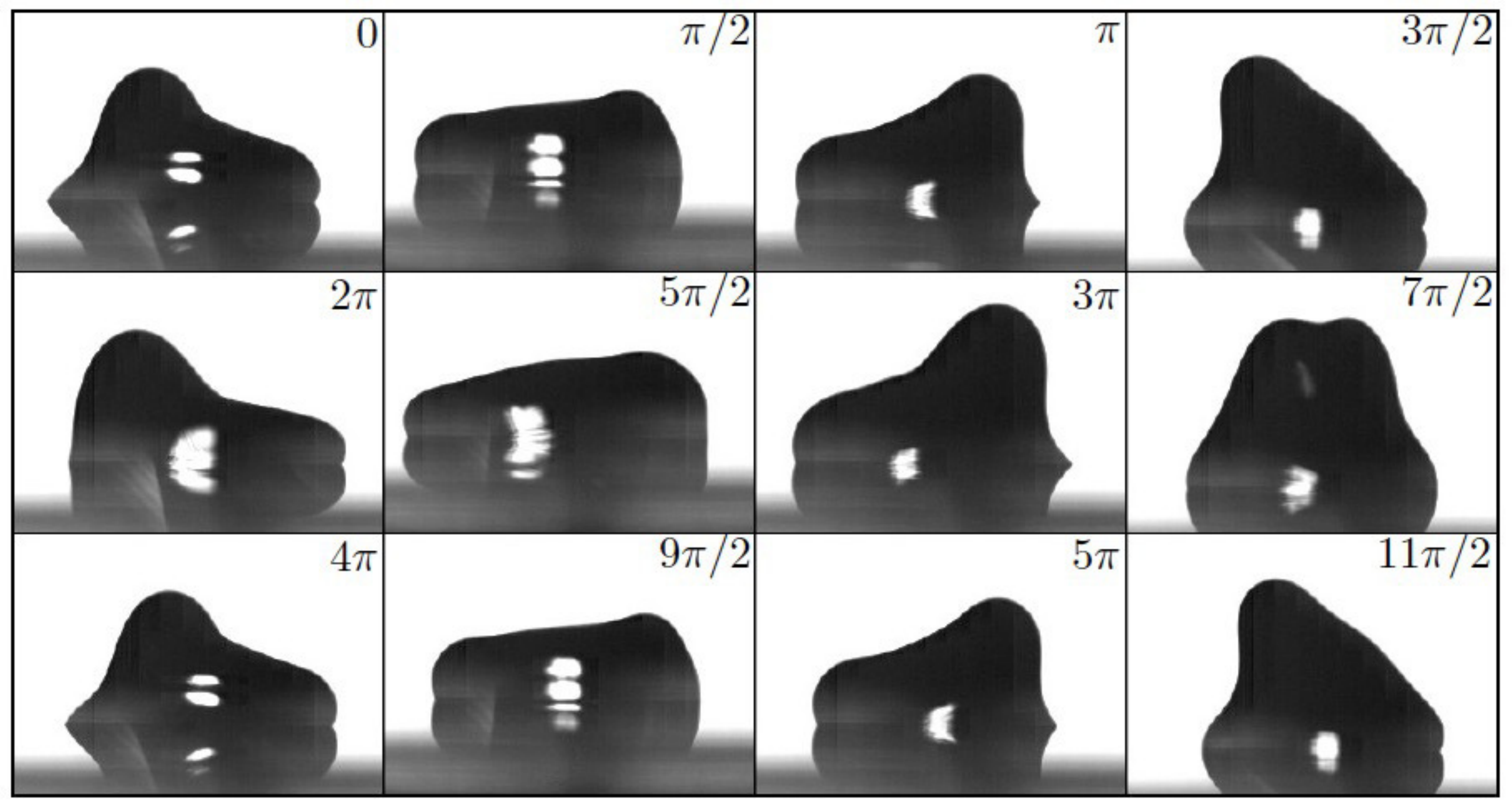}
\includegraphics[scale=0.4]{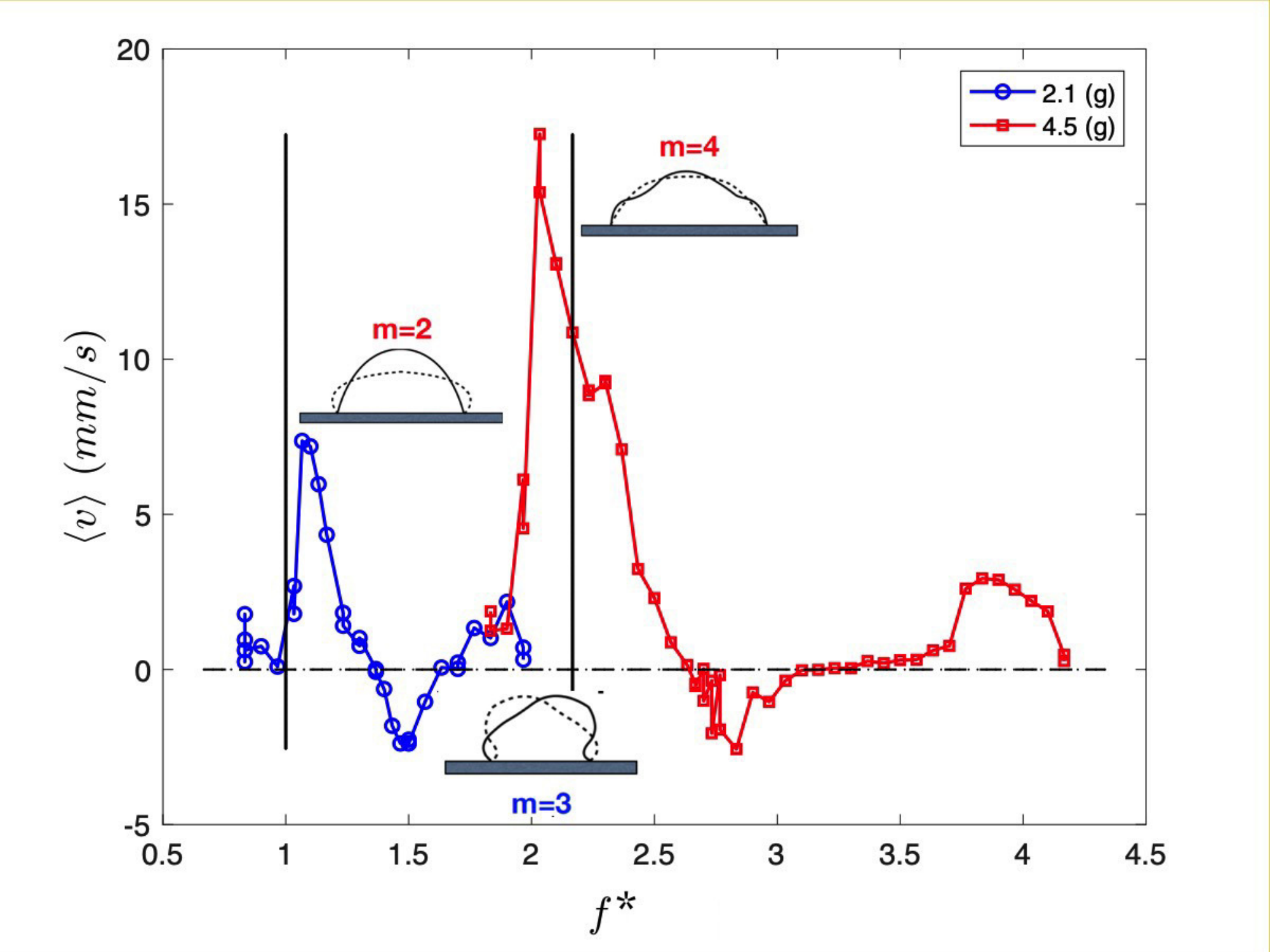}
\includegraphics[scale=0.18]{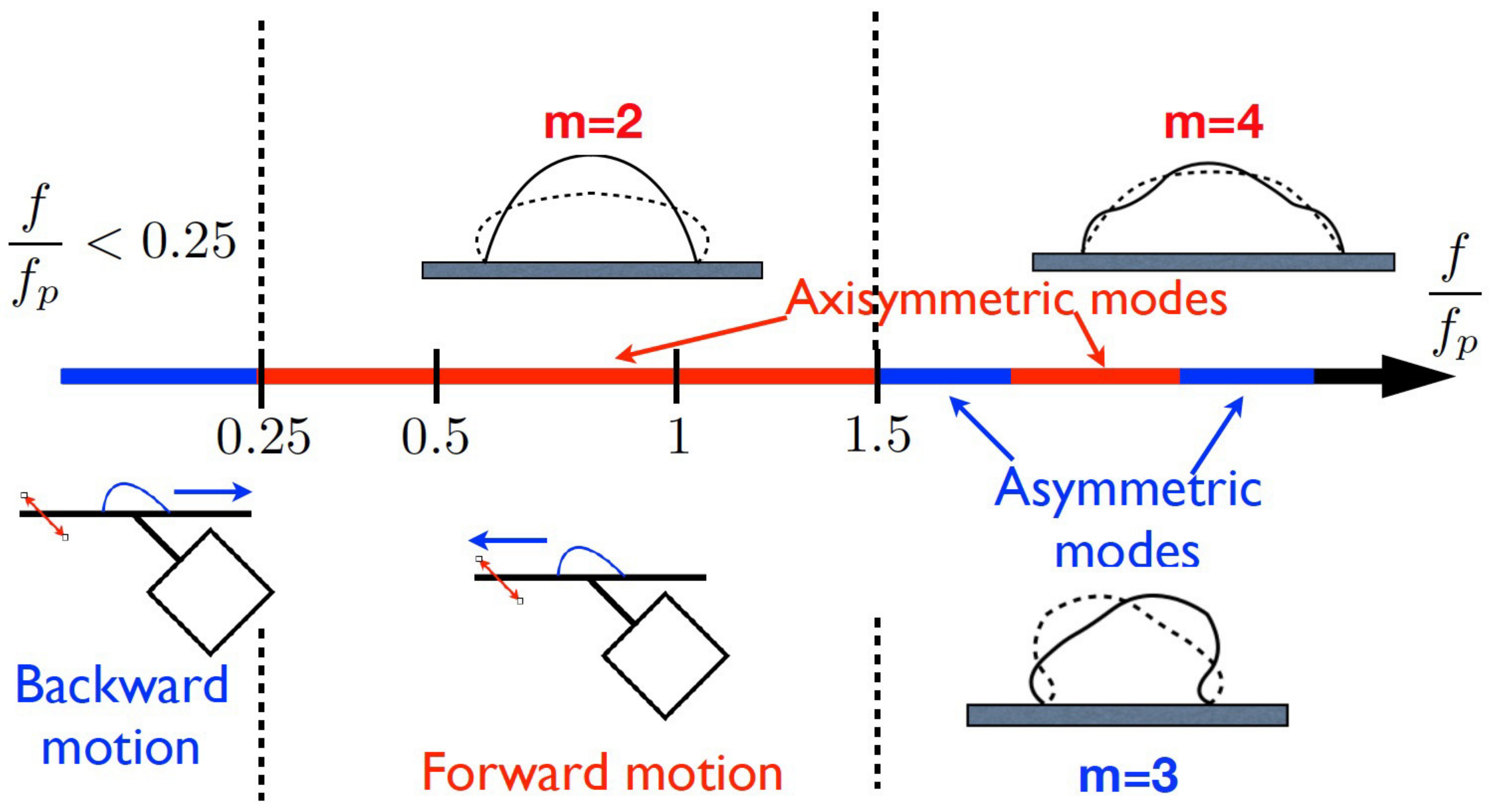}
\caption{\textit{Top} - Sequence of successive droplet shapes at relatively high forcing frequency $f^*$ = 1.48 ($f$ = 120 Hz), $V$=10 $\mu$l and low viscosity $\eta$ = 7.1 cSt. The drop experiences a backward motion (toward the right) corresponding to measurements presented in Fig.~\ref{fig:influ_freq_intro}-(b). Remarkably, the drop responds with period doubling - the shapes appear identical every two periods, induced by a parametric instability. \textit{Bottom} - Left : Drop velocity versus $f^*$ for a constant acceleration $a$ = 2.1 g and 4.5 g, with the dominant mode corresponding to each peak - Right : Simplified sketch of the overall dynamics within the whole range of $f^*$ with corresponding modes.}
\label{fig:influ_freq_intro2}
\end{figure}

A reverse motion is also observed when $f^*$ takes values around 1.5. Figure \ref{fig:influ_freq_intro}-(b) shows the drop velocity versus $A$, for $f$ = 120 Hz and $V$ = 10 $\mu$l ($f^*$ = 1.48). This backward motion appears as $A$ is set above a relatively low threshold value ($A_{\text{th}} \simeq$ 0.2 mm in the typical situation depicted here) and that contrary to the previous situation ($f^* \simeq$ 0.2), the velocity does not become positive at high $A$. Let us also notice that this high frequency backward motion is observed only for relatively low viscous liquids (here, $\eta$ = 7.1 cSt). Figure \ref{fig:influ_freq_intro2}-\textit{(Top)} shows a typical sequence of drop shapes at different phases during backward motion. Obviously, the drop deformations are mainly due to a higher order mode, but they still combine symmetric and asymmetric modes.

We showed on purpose the successive shapes during three periods of substrate oscillations : it is clear that the shape of the drop appears identical every two periods (for instance, at $\phi = \frac{\pi}{2}$ and $\phi = \frac{9 \pi}{2}$) : the drop responds at $f/2$. The period doubling results from a parametric instability. Indeed, when excited on a vibrating substrate of low CAH (weak pinning), drops experience a time-modulation of their radius, and in turn a time-modulation of their resonance frequency. This is the main ingredient for a parametric instability to occur to the drop free-surface which, in the case of vertical vibrations, leads to triplon states or star shapes \cite{Brunet_Snoeijer11,Noblin05};   

As far as we could investigate, this parametric response and the peculiar successive shapes of Fig.~\ref{fig:influ_freq_intro2}-\textit{(Top)} are necessary conditions to observe the backward motion of the drop at high $f^*$, which are satisfied only in a narrow range of $f^*$, see Fig.~\ref{fig:influ_freq_intro2}-\textit{(Bottom)}.

This strong dependence of the drop mobility on $f^*$ is revealed by measuring $<v>$ versus $f^*$ under constant acceleration $a$. Figure \ref{fig:influ_freq_intro2}-\textit{(Bottom (a) and (b))} shows the results for two values of $a$ : $a$ = 2.1 g within the range 0.8 $< f^*<$ 2 and $a$ = 4.5 g within the range 1.8 $< f^* <$ 4.2. The reason for taking two values for $a$, is that $<v>$ has too small values for $a$ = 2.1 g. and $f^*>2$. These measurements essentially illustrate how strong the drop velocity can depend on $f^*$, and how it is related to the excitation of the different modes. The vertical plain and dashed lines point out the resonance frequencies of the modes (1) to (4), the rocking mode of lower frequency is not represented here. Thus it is striking how these resonances are related to the maxima of velocity with $f^*$ in either the normal or inverse directions.

\subsubsection{Quantitative study in a model situation}

\begin{figure}
\subfigure[]{\includegraphics[scale=0.41]{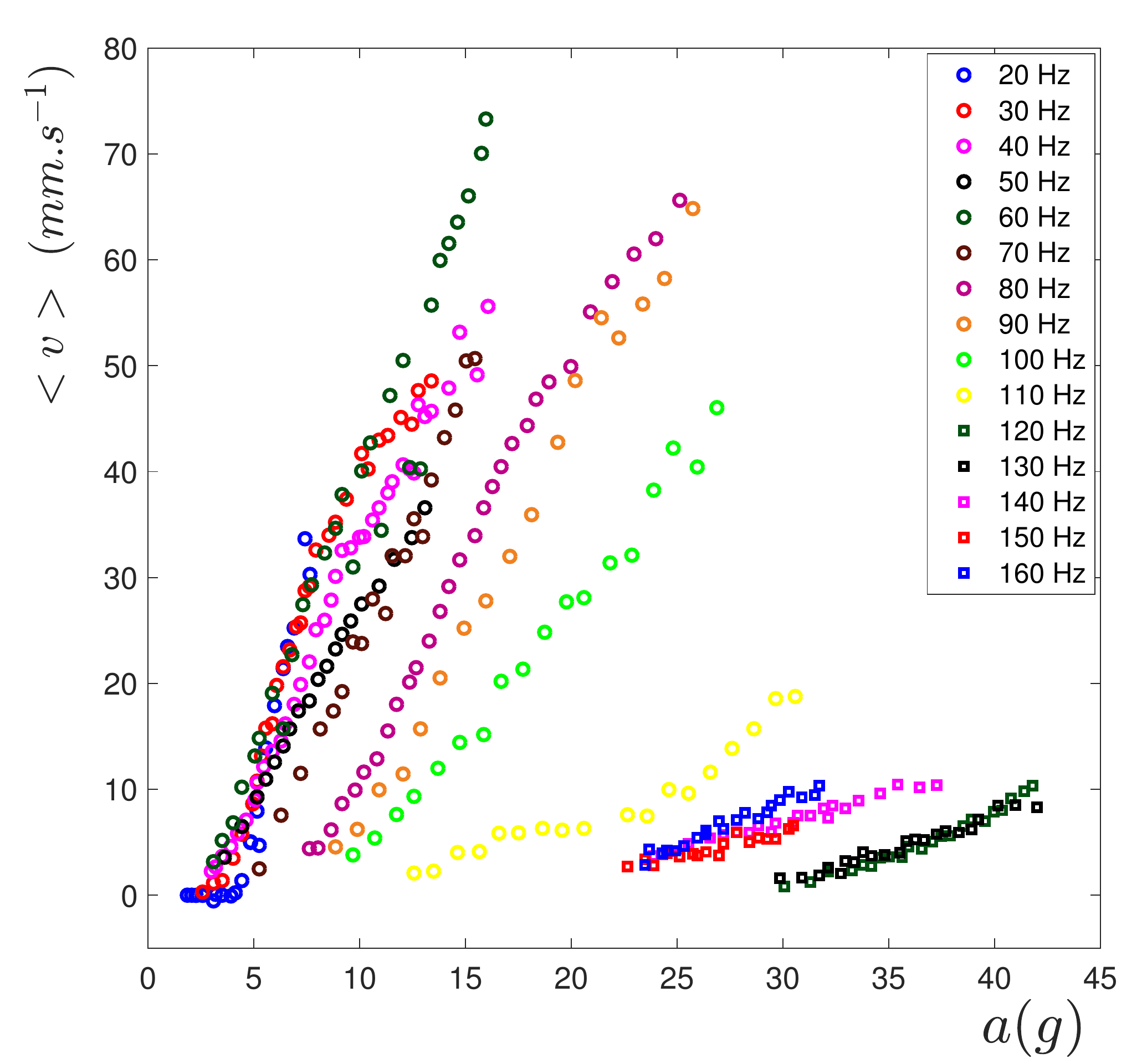}}
\subfigure[]{\includegraphics[scale=0.41]{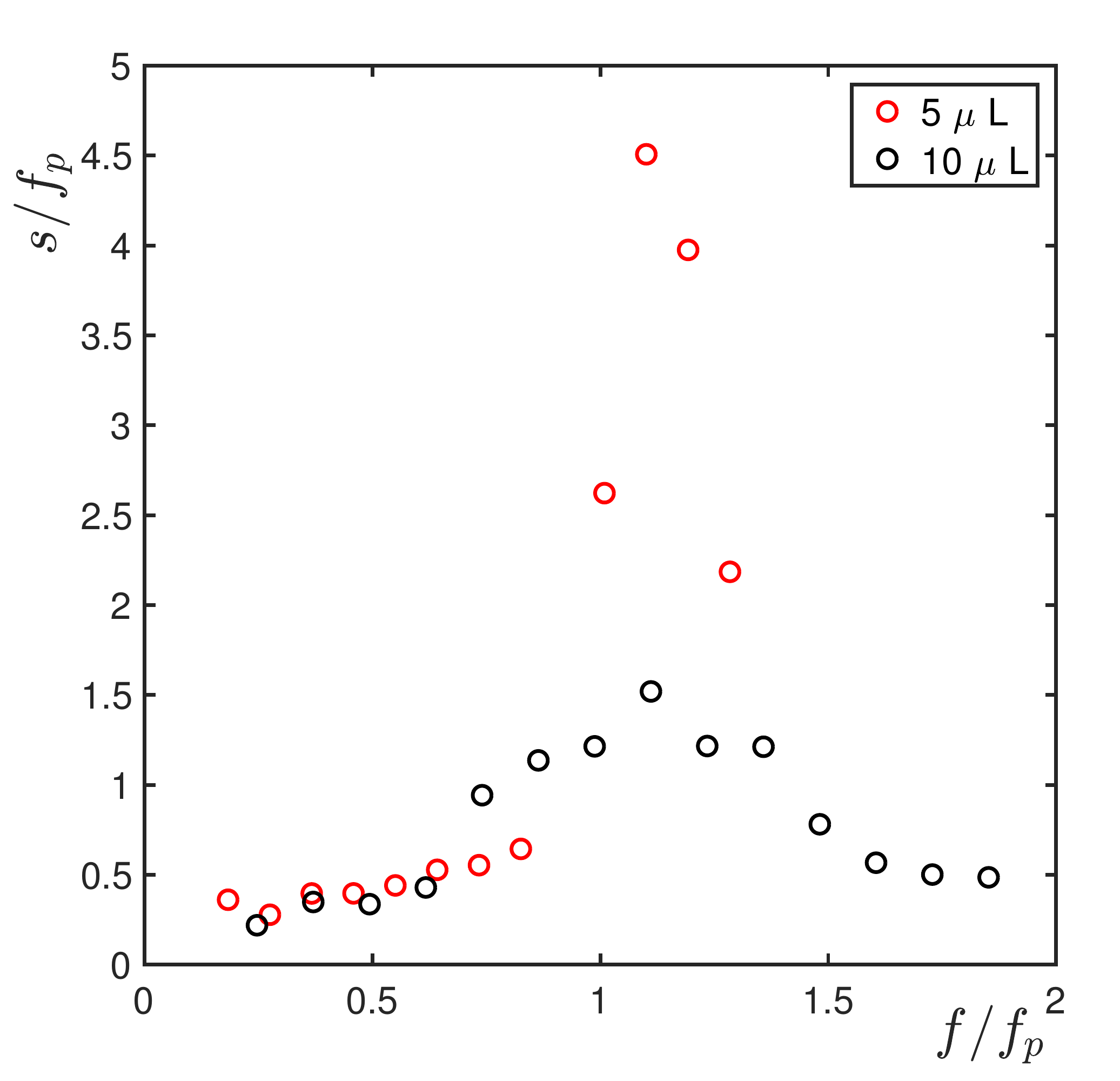}}
\subfigure[]{\includegraphics[scale=0.41]{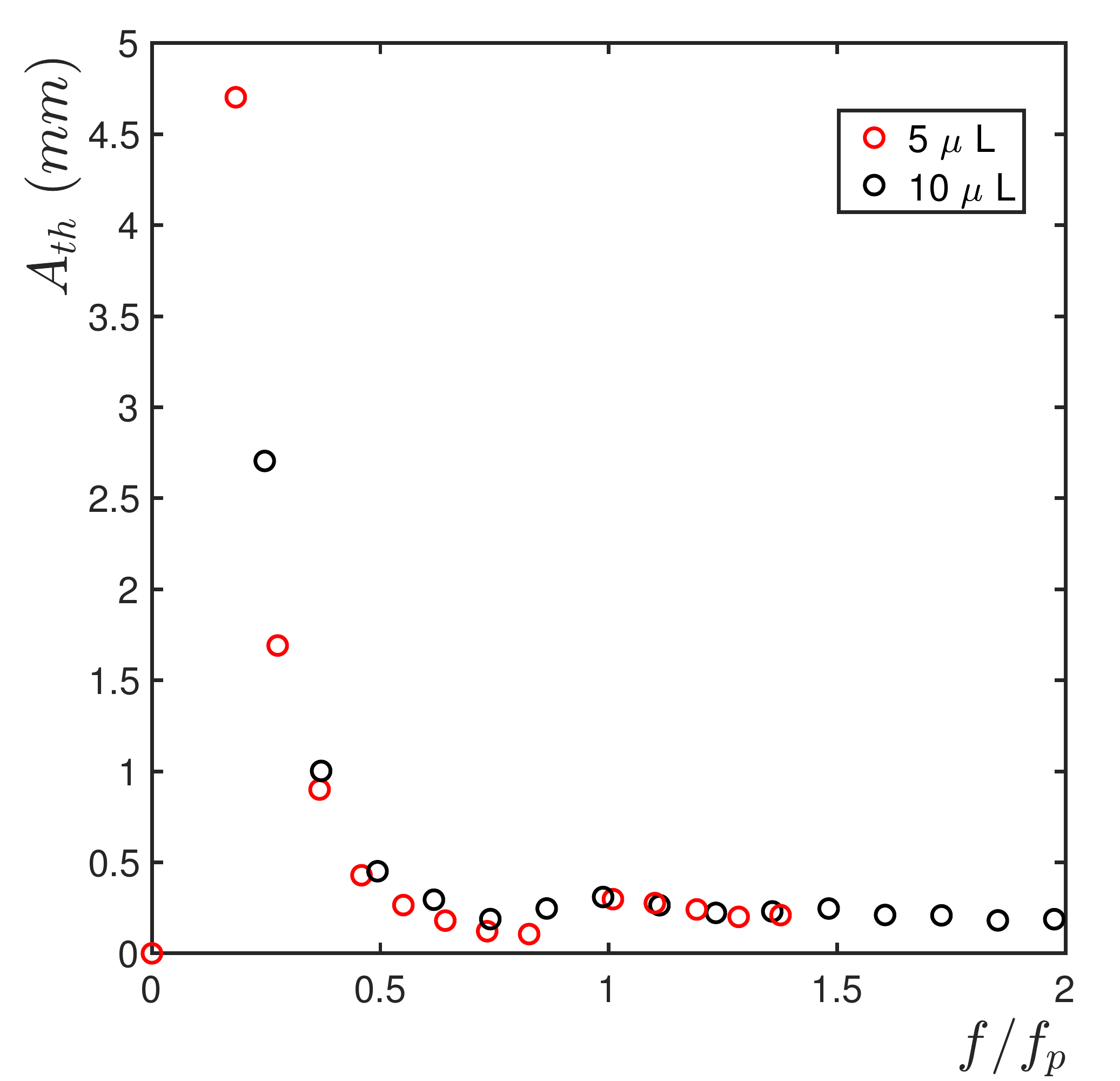}}
\subfigure[]{\includegraphics[scale=0.41]{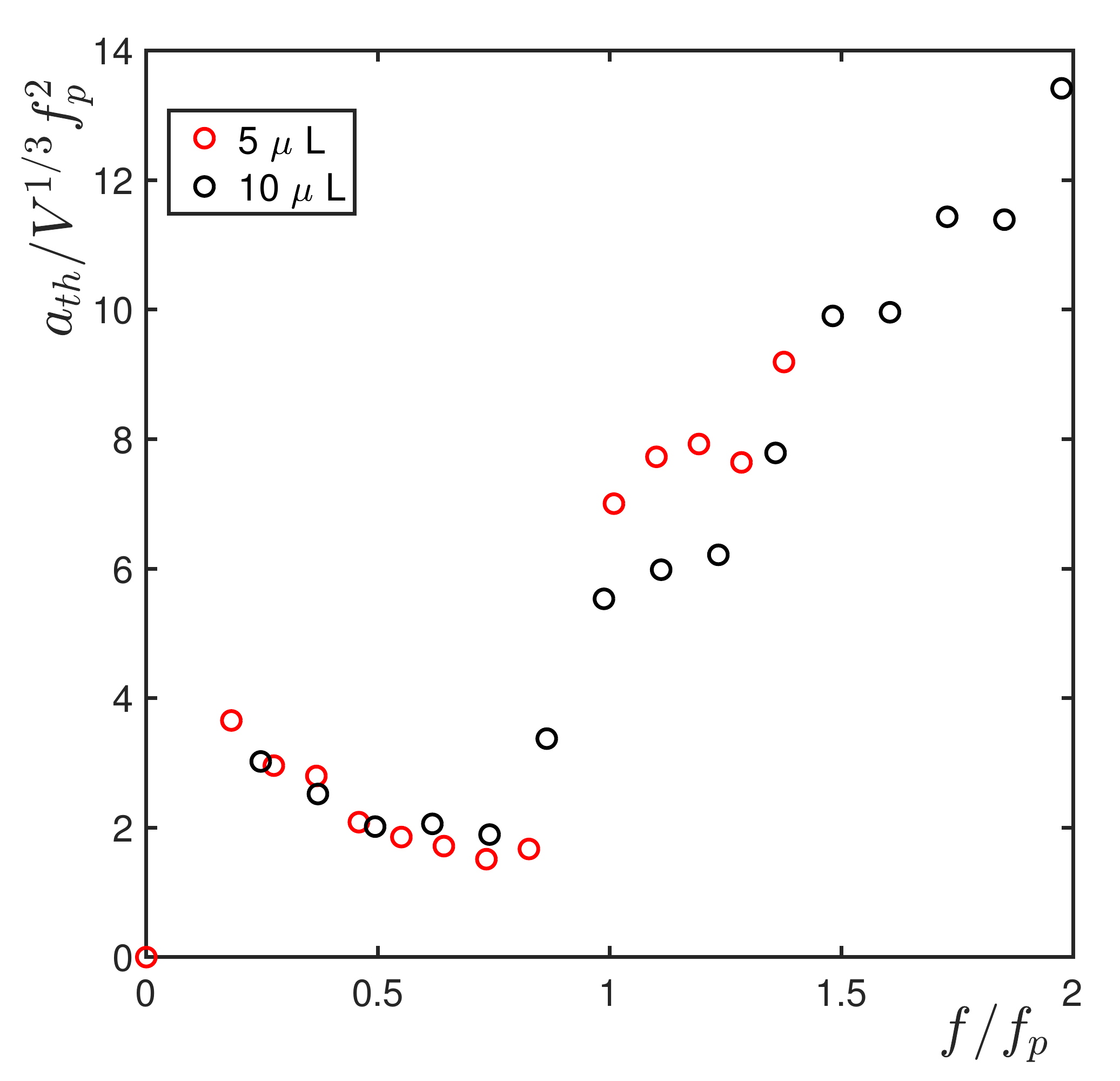}}
\caption{(a) Velocity of the drop center of mass versus acceleration $a = A \omega^2$, for different frequencies $f$ from 20 to 160 Hz. Drop volume $V$= 10 $\mu$l and viscosity $\eta$ = 28.8 cSt. Angle $\alpha$ = 60$^{\circ}$. (b) Dimensionless mobility coefficient for $V$ = 5 and 10 $\mu$l, same conditions as (a) otherwise. (c) The amplitude threshold versus $\frac{f}{f_p}$ for $V$ = 5 and 10 $\mu$l. (d) The acceleration threshold $a_{\text{th}}$ renormalized by $V^{1/3} f_p^2$, versus $\frac{f}{f_p}$ for $V$ = 5 and 10 $\mu$l.}
\label{fig:influ_freq}
\end{figure}

We now focus on the more common forward motion in the range of $f^*$ roughly between 0.2 and 1.8. We quantitatively investigate the influence of $f$ for this model situation, where only positive values of $<v>$ are obtained. Figure \ref{fig:influ_freq}-(a) shows $<v>$ versus $a$, for different values of frequency $f$, from 20 to 160 Hz. The angle $\alpha$ is kept at 60$^{\circ}$. These results suggest that data can mostly be well fitted by taking the equation (\ref{eq:vmoy}) with $\chi$ = 1.

In Figure \ref{fig:influ_freq}-(b), the dimensionless mobility $\frac{s}{f_p}$ is plotted versus $\frac{f}{f_p}$. Let us mention that when $<v>$ slightly deviates from a linear dependence with $A$ and $a$, for instance at strong amplitude $A$ for $f$= 30 Hz or 110 Hz, only the data at moderate $A$ are considered for the determination of $s$. These experimental results suggest an optimum of mobility for $\frac{f}{f_p} \simeq$ 1.1 for both the tested volumes (it corresponds roughly to $f$ = 90 Hz at $V$=10$\mu$l). This maximum is much sharper for $V$ = 5 $\mu$l than for $V$ = 10 $\mu$l. Even considering that $f_p$ is lower for 5 $\mu$l than for 10 $\mu$l by a factor of $\sqrt{2}$, it turns out that smaller drops have a much sharper dependence on frequency around $f_p$. The value of $f_R$ is about half that of $f_p$.

Then, we extract the threshold $A_{\text{th}}$ versus $f^*$, that is plotted in Fig.~\ref{fig:influ_freq}-(c) for $V$ = 5 and 10 $\mu$l. We also plot the threshold in acceleration $a_{\text{th}}$, divided by a characteristic acceleration $V^{1/3} f_p^2$ build on the drop size, versus $f^*$. The values of $A_{\text{th}}$ correspond to the threshold for forward motion. When $f^* <$ 0.25, as seen in Fig.~\ref{fig:influ_freq_intro}, this forward motion is preceded by a backward motion of smaller velocity at $A < A_{\text{th}}$. The backward motion can exist within a relatively large range of $A$, considering the large values of $A_{\text{th}}$ in this range of low $f^*$. Let us remark that the backward motion at larger $f^*$ ($\simeq$1.5) is not observed here, as we chose a value of $\eta$ (=28.8 cSt) large enough to dismiss the parametric forcing shown in Fig.~\ref{fig:influ_freq_intro2}-{Top}.

While $A_{\text{th}}$ variations are essentially within the range $f^* <$ 0.5 (or $f < f_r$), converging to a constant small value at higher frequency, $a_{\text{th}}$ experiences a sharp increase above a value of $f$ between $f_r$ and $f_p$ (corresponding to $f^* \simeq$ 0.75), where it reaches its minimal value Fig.~\ref{fig:influ_freq}-(d). To make the two data sets for both volumes to collapse with each other, $A_{\text{th}}$ is plotted versus $f/f_p$. Such a rescalling is obtained for $a_{\text{th}}$, by dividing it by the characteristic acceleration $V^{1/3} f_p^2$.

If we briefly come back to the question of the determination of the exponent $\chi$, given the aforementioned discrepancy between existing experiments and models (some of them indeed predict this linear dependence while others predict $\chi$ between 1.5 and 2), we can conclude here that the averaged velocity $<v>$ generally follows a linear relationship ($\chi$=1) with forcing amplitude $A$ (or acceleration $a$) for fixed $f^*$ and $\delta^*$. Still, some of our measurements showed possible larger values for $\chi$ (slightly higher than 2, see Fig. 1 in \textit{Supplementary Materials}) in a narrow range of $f^*$ around 1.5. \textcolor{blue}{This is clearly different from quadratic laws predicted by theoretical and numerical studies \cite{Benilov10,Benilov11,Benilov13,Bradshaw_Billingham16,Bradshaw_Billingham18}. Though, we do not have clear explanation for this non-intuitive behavior.}

\subsection{The influence of viscosity}

\begin{figure}
\subfigure[]{\includegraphics[scale=0.55]{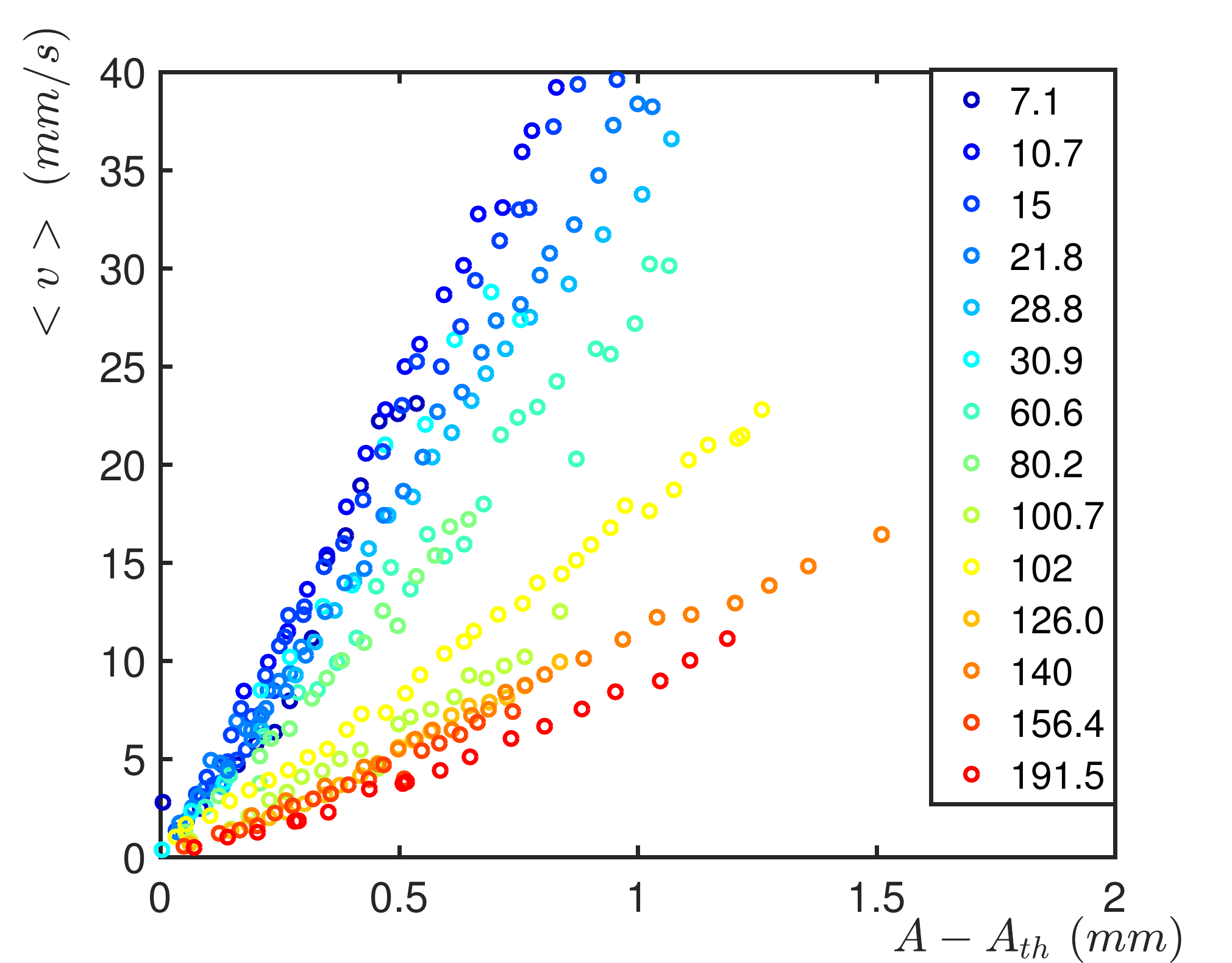}} \\
\subfigure[]{\includegraphics[scale=0.435]{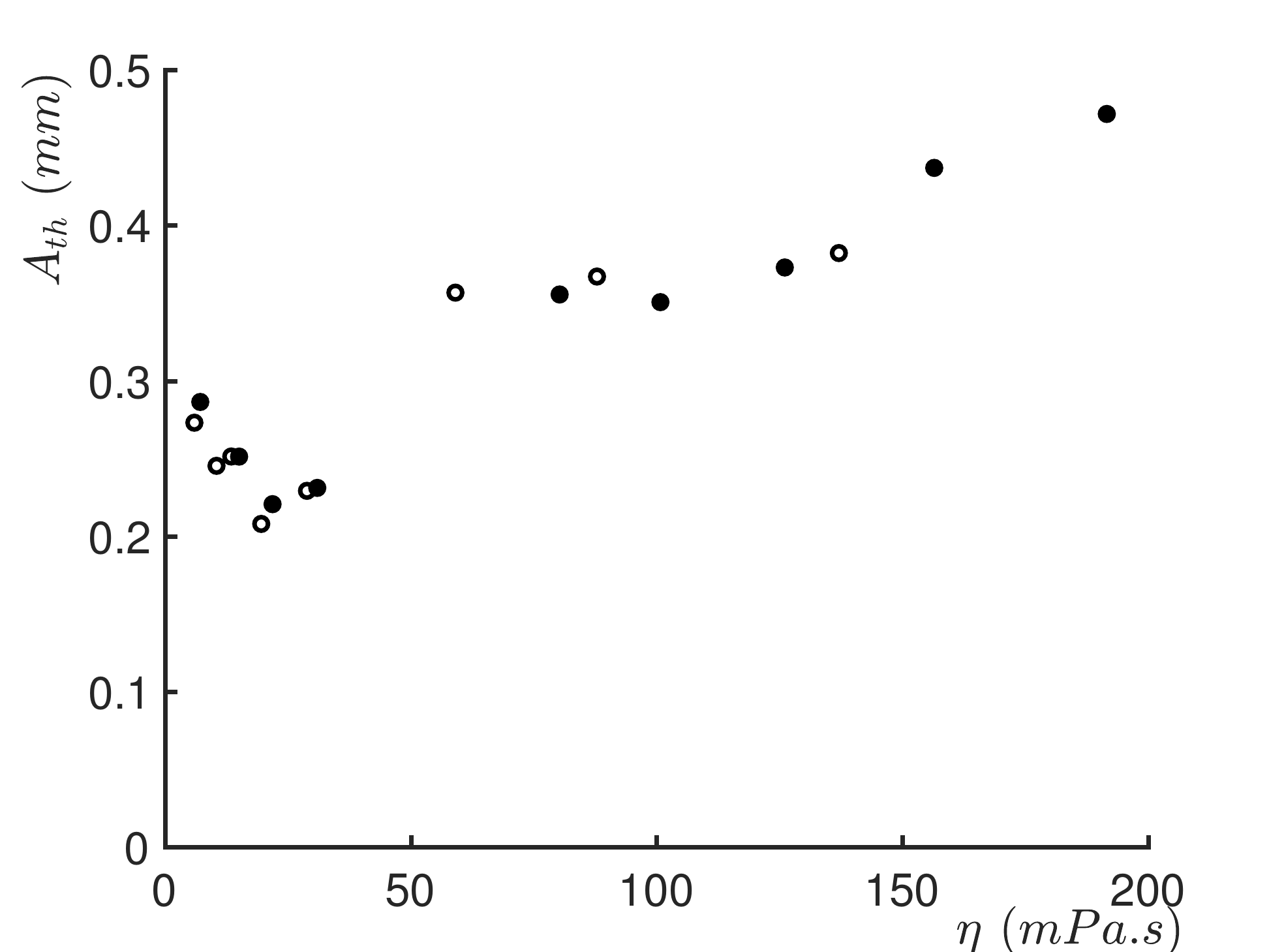}}
\subfigure[]{\includegraphics[scale=0.435]{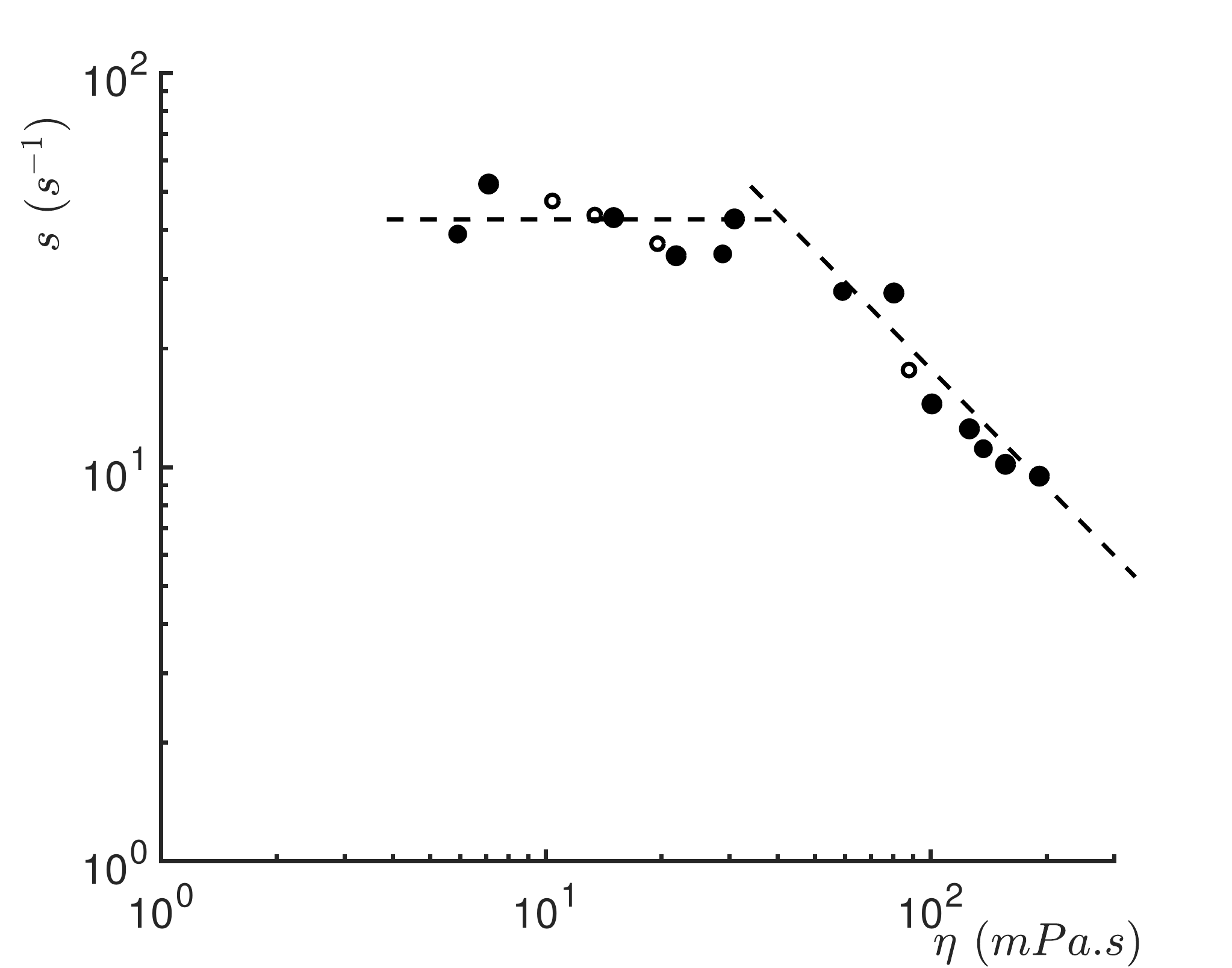}}
\caption{(a) Average velocity of the drop center of mass versus amplitude of vibration, for different angles $\alpha$ between the horizontal and the axis of vibrations. $f$ = 50 Hz, $V$ = 10 $\mu$l. Extracted from these results, (b) the amplitude threshold versus dynamic viscosity, and (c) the mobility coefficient $s$ versus viscosity in log-log plot.}
\label{fig:influ_visco}
\end{figure}

We now present quantitative results of the dependence of the average velocity $<v>$ on the viscosity $\eta$, keeping $V$ and $f$ constant respectively at 10 $\mu$l and 50 Hz ($f^*$=0.67). Let us first note that liquid viscosity has little influence on the shape and resonance frequency of the drop eigenmodes - although it influences the amplitude of deformations. Viscous shear is one of the main source of the friction force experienced by a drop moving on a substrate, both along the liquid-solid interface and near the contact-line. Furthermore, viscosity is supposed to influence the thickness of the BL $\delta$ - then the ratio $\delta^*$ of liquid thickness that responds in phase with the substrate vibrations, see eq.~(\ref{eq:boundarylayer}). 

Figure \ref{fig:influ_visco}-(a) presents $<v>$ versus $(A-A_{\text{th}})$ for several values of $\eta$, varying from 5.9 to 191.5 cSt. 

The threshold $A_{\text{th}}$ is plotted versus $\eta$ in Figure \ref{fig:influ_visco}-(b). The dependence of $A_{\text{th}}$ on $\eta$ is very weak, and only a slight increase with $\eta$ can be noticed above 50 cSt. 

It turns out that $\eta$ has a significant influence on the drop mobility through the coefficient $s$. As previously, $<v>$ increases linearly with $A$, so that we keep $\chi$=1 in eq.~(\ref{eq:vmoy}), see Fig.~\ref{fig:influ_visco}-(a). The general trend is that a higher $\eta$ slows down the drop for the same forcing ($f$ and $A$ or $a$). At 50 Hz, $\delta$ ranges from 0.486 mm (for 5.9 cSt) to 2.76 mm (for 191.5 cSt), and taking the drop shape as a hemispherical cap, $h = R = \left(\frac{3 V}{2 \pi}\right)^{1/3}$ and $V$ = 10 $\mu$l, $\delta^*$ ranges from 0.29 (for 5.9 cSt) to 1.64 (for 191.5 cSt). 

Figure \ref{fig:influ_visco}-(c) shows a decrease of the mobility $s$ with viscosity, but only in the higher range of $\eta$. At a simplified level of description, one could state that within a range of low viscosity, $s$ is almost constant with $\eta$ (with only a slight decrease). The crossover occurs around 50 cSt, hence for $\delta_c \simeq$ 1.41 mm, or $\delta^*_c \simeq$ 0.83.

We propose a qualitative explanation for these two distinct behaviors. At relatively high viscosity, the dissipation occurs in the whole volume (within a liquid height $h$). The order of magnitude of the momentum per unit volume relative to the forcing term reads scales as $\frac{\rho \mathcal{A}^2 \omega^2}{h}$, while the order of magnitude of viscous shear opposing the forcing scales as $ \eta \frac{v}{h^2}$, so that the balance between the two terms leads to an order of magnitude for the averaged velocity :

\begin{equation}
\eta \frac{<v>}{h} \sim \rho \mathcal{A}^2 \omega^2
\label{eq:scaling_highvisco}
\end{equation}

\noindent hence, an averaged velocity that scales with the inverse of $\eta$. Conversely at relatively low viscosity, the dissipation only takes place in a layer of thickness $\delta < h$ and reads $ \eta \frac{<v>}{\delta^2}$. Therefore, substituting the expression of $\delta$ it yields :

\begin{equation}
<v> \sim \frac{\mathcal{A}^2 \omega}{h}
\label{eq:scaling_lowvisco}
\end{equation}

\noindent hence an averaged velocity independent on viscosity. Of course, this reasoning remains qualitative, as this does not take into account the influence of the frequency emphasized in Figs.~\ref{fig:influ_freq}, which involves the eigenmodes of the drop. In this sense, the amplitude $\mathcal{A}$ here should be considered as the amplitude of the drop oscillations (taken at the centre of mass or at the free-surface) rather than the forcing amplitude $A$, and should include a dependence on $f$ like :  $\mathcal{A} \sim A^{\xi} \mathcal{F}(f)$, where $\xi$ is a positive exponent. The scaling of $<v> \sim A^{\chi}$ , with experiments showing $\chi \simeq$ 1, tends to suggest that $\xi$ should roughly equal $\frac{1}{2}$. Finally, the dissipation at the contact-line is not included in this qualitative reasoning, given the relatively low CAH resulting from the SAM coating. 

\section{DISCUSSIONS}

\subsection{Comparisons with existing models}

Our measurements carried out over a large span of $\alpha$, $f$ and $\eta$, show that when subjected to slanted vibrations, sessile droplets \textit{always} experience a directional motion providing the forcing amplitude is strong enough. In most situations, the motion is oriented toward the direction that corresponds to that of the substrate displacement during its upward-moving phase (i.e. toward the left in the configuration of Fig.~\ref{fig:extract_drop}). Incidentally, this corresponds to the climbing motion observed in \cite{Brunet07,Brunet09} for vertical vibrations of a tilted substrate. 

In the light of these results, we can come back to the first question stated at the end of the introduction, namely the influence of the relative importance of viscosity and inertia in the drop's inner flow, which can be quantified by $\delta^*$, and its influence on the mobility (optimal conditions). A related question is that of the 'minimal ingredients' to get directional motion at relatively moderate forcing. 

Measurements at high viscosity, i.e. $\delta/ h >$ 1, show that when the drop responds quasi-statically and in phase to the forcing, directional motion remains possible above a forcing threshold comparable to threshold values measured for much lower viscosities, although the drop velocities remain relatively small. These measurements confirm the results by John and Thiele \cite{John_Thiele10}, who showed that the minimal ingredient for directional motion relied on the successive flattening and stretching of the shape over one period, while the drop rocks left and right, which remains true for a viscous drop. These deformations then lead to a non-linear mobility and anharmonic response for the drop. 

On the other side, the recent model with inviscid drops by Bradshaw and Billingham \cite{Bradshaw_Billingham18} includes dissipation through contact-angle hysteresis and more complex laws for dynamical wetting that involve multiple-valued contact-line velocities versus contact-angle. These studies evidenced an optimal value of CAH at roughly 5$^{\circ}$, a feature which was experimentally observed for sessile droplets displaced by asymmetric vibrations \cite{Mettu_Chaudhury11}, although for a slightly larger optimal CAH value. The other approach described in \cite{Bradshaw_Billingham16}, also by Bradshaw and Billingham, showed that neglecting both viscous and inertial effects still enables directional motion. Hence, this quasi-static equilibrium between gravity and surface-tension forces, although in a situation difficult to reproduce in experiments, could also constitute a situation with 'minimal ingredients'  for directional motion. 

About the question on a value of $\delta^*$ for optimal mobility, our experiments are only partly conclusive. The results of Fig.~\ref{fig:influ_visco}-(c) show that $s$ remains almost constant below $\eta$=50 cSt ($\delta^*_c$ = 0.83), which suggests that for a fixed value of $A$, the dimensionless velocity $<Ca> = \frac{\eta <v>}{\gamma}$ is maximal around $\delta^*_c$. The averaged motile force per unit length being proportional to the product $\eta <v>$, this motile force seems to reach a maximum near $\delta^*_c$, suggesting that a subtle balance in the relative importance of inertia and viscosity is required for this optimum.

As shown experimentally by Noblin \textit{et al.} \cite{Noblin09}, the phase-lag between the pumping and rocking modes response can be tuned in order to obtain optimal mobility. In our experiments, this phase-shift should depend on both the dimensionless quantities $f^*$ and $\delta^*$, in a non-trivial way. However, in the present situation, the phase-lag cannot be controlled. The influence of this phase-shift on the optimum of mobility is here suggested by the dependence of the mobility $s/ f_p$ on the reduced frequency $f^*$ (Figure \ref{fig:influ_freq}-(b)), where the optimal was found for $f^*=$1 and which was also found for a nearby value in \cite{Bradshaw_Billingham18} : this situation optimises the response of the pumping mode over the rocking one. Still, we can attempt to address this point in more details by having a deeper look on the dynamics of the contact-lines ($x_b (t)$ and $x_f (t)$) and its relation to the time-evolution of the dynamical CAs ($\theta_b (t)$ and $\theta_f(t)$).

\subsection{Relationship between global motion and unsteady contact-line dynamics}

\begin{figure}
\subfigure[]{\includegraphics[scale=0.42]{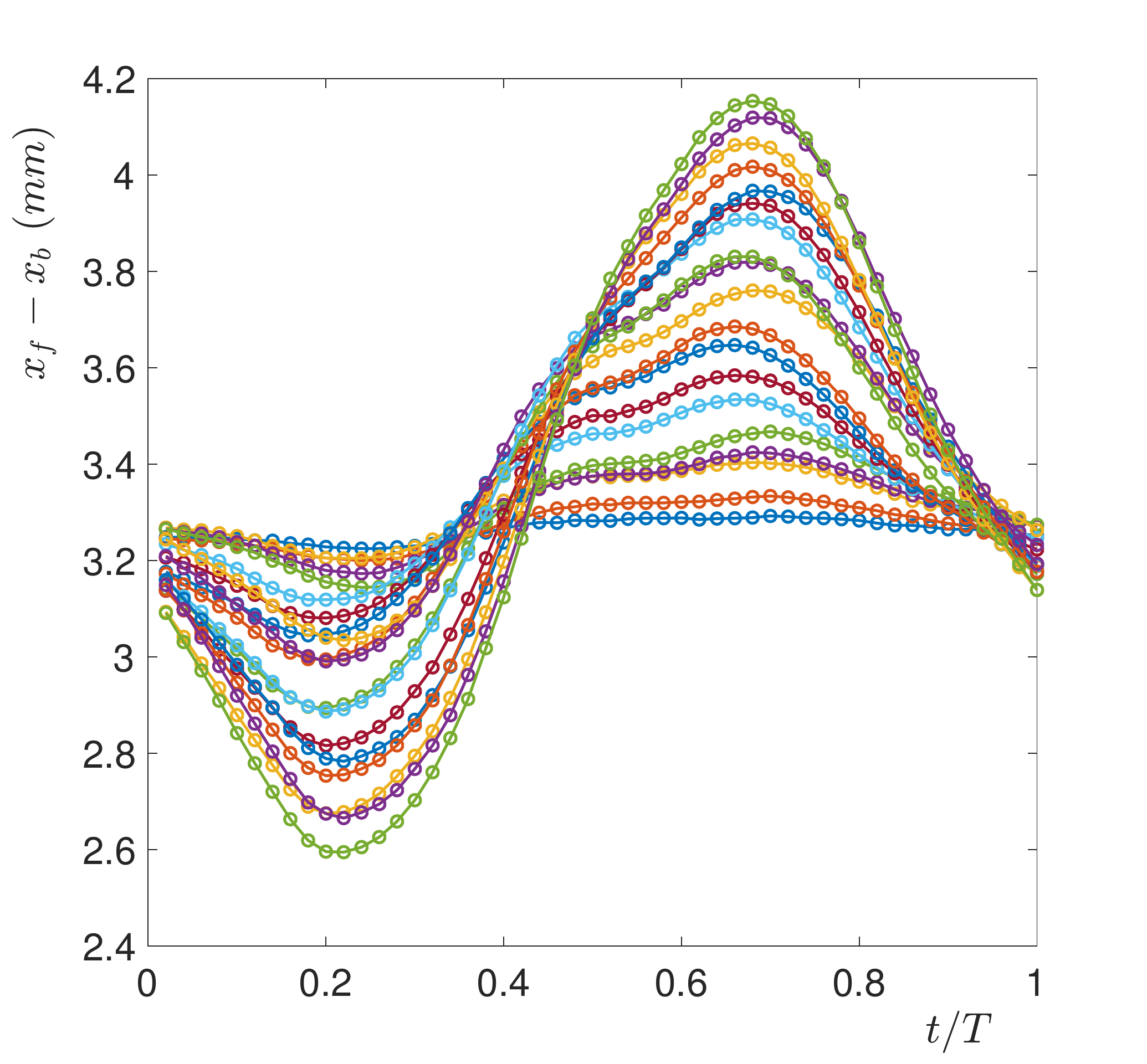}}
\subfigure[]{\includegraphics[scale=0.42]{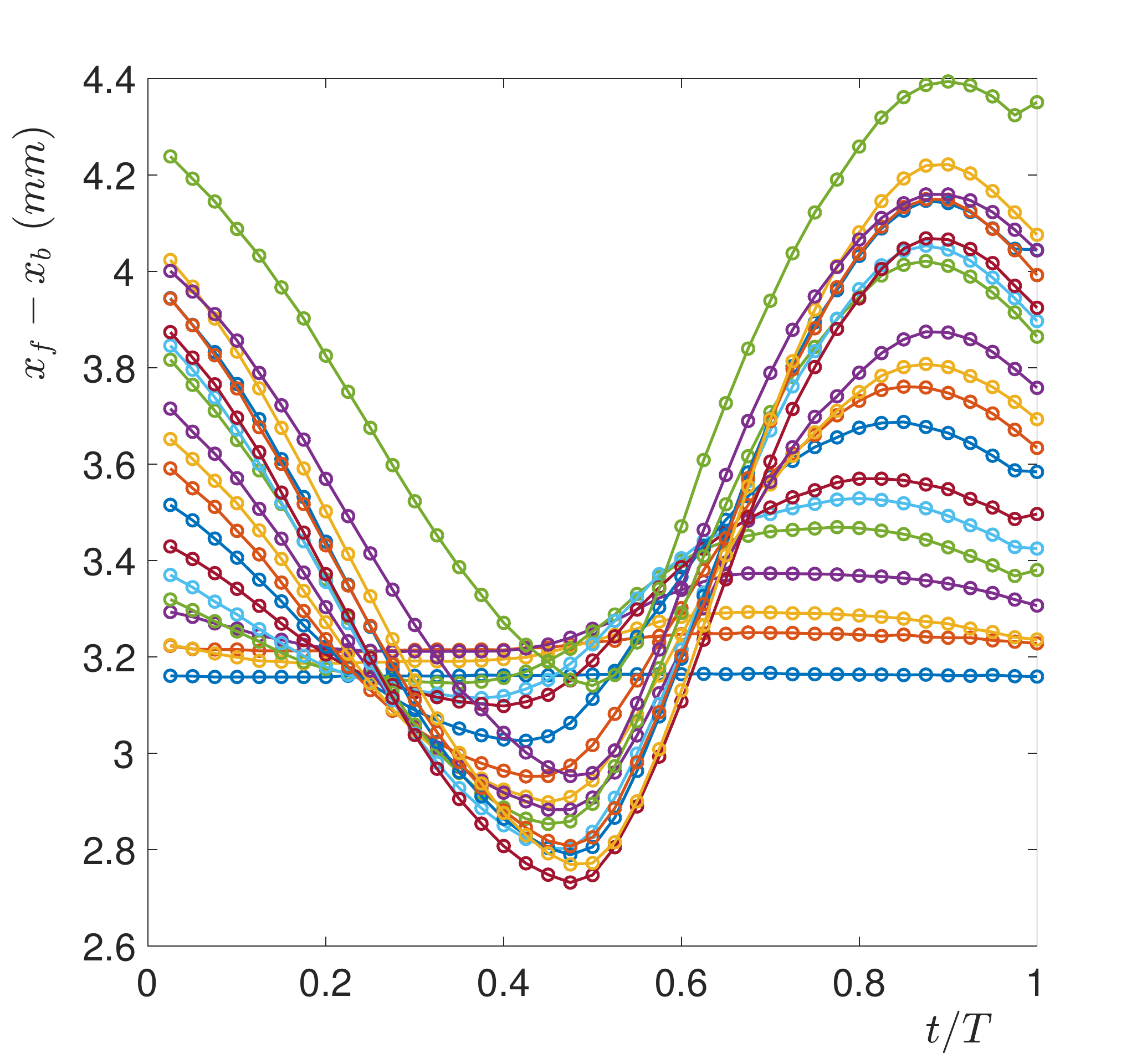}}
\subfigure[]{\includegraphics[scale=0.42]{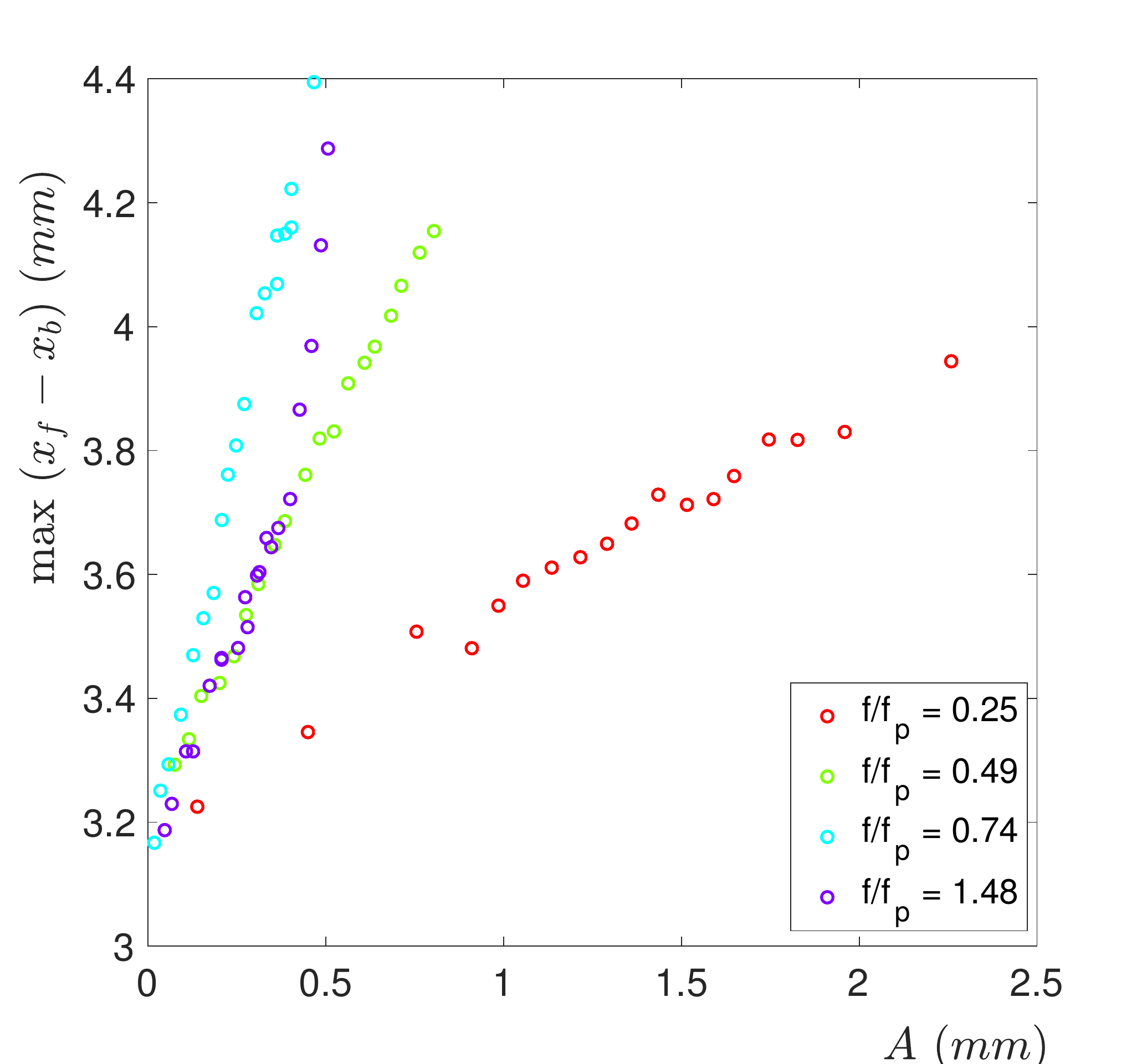}}
\subfigure[]{\includegraphics[scale=0.42]{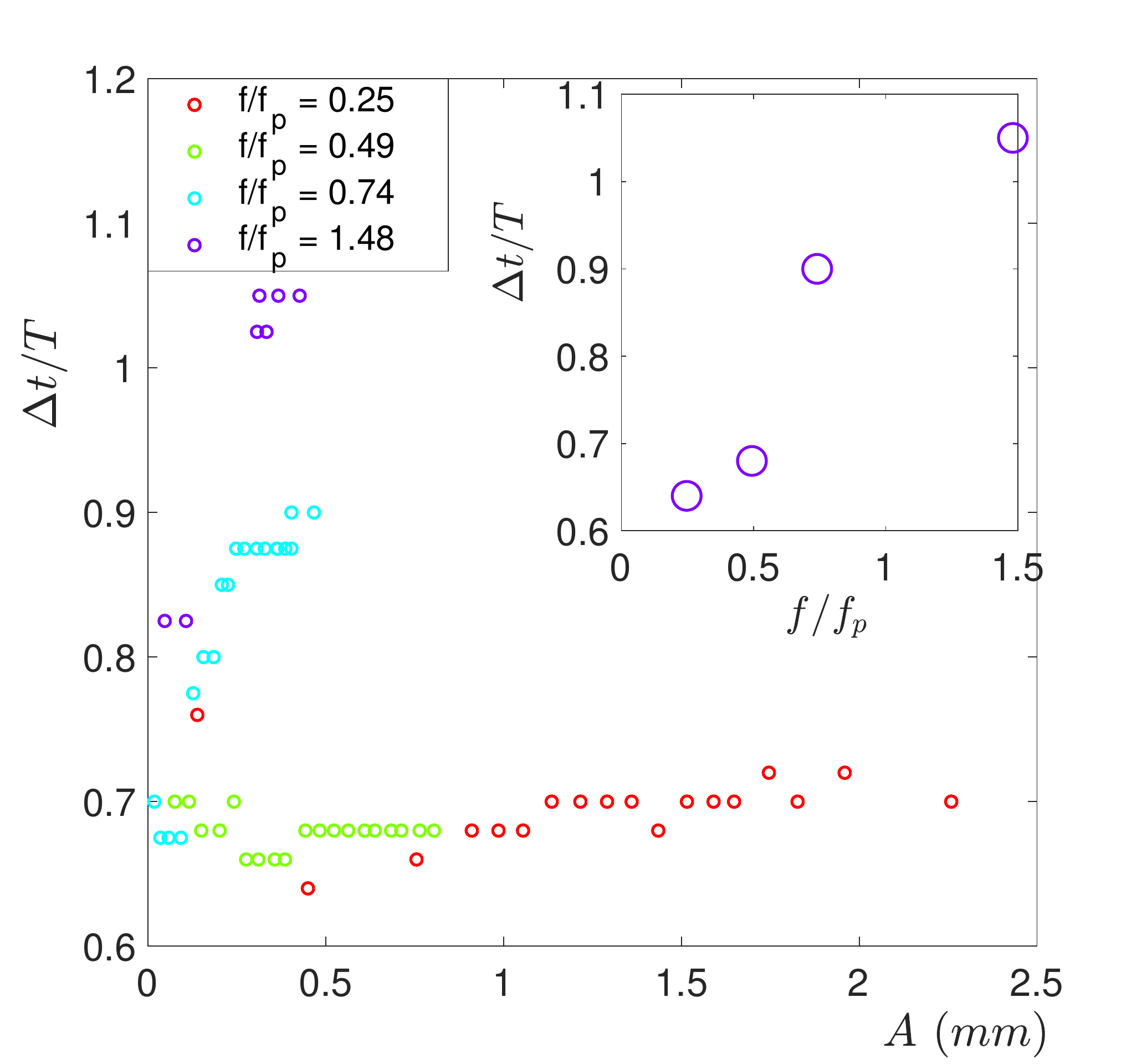}}
\caption{(a, b) Time-evolution of the basal diameter $(x_f - x_b)$, over one period and for various forcing amplitude $A$. $V$ = 10 $\mu$l and $\eta$ = 7.1 cSt, and (a) $f^*$ = 0.49, (b) $f^*$ = 0.78. The time $t$=0 corresponds to the maximal position of the platform. (c) Maximal basal diameter versus $A$, for different $f^*$, (d) Phase-shift (expressed as dimensionless time-lag) between the basal diameter response and the forcing vibration, versus $A$, for different $f^*$. Insert : Time-lag versus $f^*$ for $A$=0.4 mm.}
\label{fig:basaldiameter}
\end{figure}

What is clear from a coarse observation of the time-evolution of $x_b (t)$ and $x_f (t)$ (Fig.~\ref{fig:setup_vibrations2}-Bottom), is that these two positions generally vary with distinct amplitude and phase-shift (with respect to the forcing vibration). This is a direct consequence of the aforementioned phase-shift between the pumping and rocking modes.

When one attempts to be more quantitative regarding the measured phase-lag between contact-line velocity and contact-angle time-evolutions, let us recall that experiments by Sartori \textit{et al.} \cite{Sartori15} did not evidence any obvious correlation with the drop motion. This absence of clear trend was also noticed in our own experiments, carried out at different viscosity and various $f^*$, as described in more details in \textit{Supplementary Materials} (Figs. 2).

The only fact that remains recurrent in most of the aforementioned experimental studies, is that the relationship between contact-line velocity and dynamical CAs shows complex behaviour taking non-singled values \cite{Brunet07,Noblin04,Perlin95,JiangSchutzPerlin04,Xia_SteenJFM18}. This complexity was included in numerical models \cite{Bradshaw_Billingham18} and was shown to increase the efficiency of the directional motion, particularly in the presence of CAH. This complex behavior is generally out of the scope of usual contact-line hydrodynamical theories \cite{Dussan,degennes85,Voinov,Eggers09}. 

Still, we attempted to grasp a quantitative measurement of the pumping mode response, via the time-evolution of the basal diameter, namely $(x_f - x_b)$.  Figs.~\ref{fig:basaldiameter}-(a,b) show the results for two values of $f^*$ and different forcing amplitude $A$. As expected, the peak-to-peak variations of $(x_f - x_b)$ grow with $A$. The time $t$=0 corresponds to the maximal vertical position of the vibrating bench, hence of $A$. Clearly, the phase shift, extracted from the time when $(x_f - x_b)$ is maximal, shows significant change with $f^*$, while the dependence with $A$ remains weak. Figure \ref{fig:basaldiameter}-(c) shows the maximal value of $(x_f - x_b)$ with $A$, and figure \ref{fig:basaldiameter}-(d) shows the phase-shift (as a dimensionless time-lag $t/T$) between $(x_f - x_b)$ and the forcing vibrations, versus $A$. Let us note that for $f/f_p$ = 1.48, the basal diameter showed period-halving and hence the phase-shift could not be simply determined, as $(x_f - x_b) (t)$ can exhibit two distinct maxima : one is at $t/T >$1 (the one which is here plotted) and the other close to $t/T$=0. This situation corresponds to a drop with backward motion, see Fig.~\ref{fig:influ_freq_intro}-(b) and Fig.~\ref{fig:influ_freq_intro2}-(a).

While for $f^* \ll$ 1, the time-lag remains between 0.6 and 0.7 - hence, slightly later than a phase-opposition, it almost reaches one (almost in phase) when $f^* \simeq$ 1, hence close to the optimal of mobility found in Fig.~\ref{fig:influ_freq}-(b). Therefore, to the best of our knowledge, the optimal of mobility seems to appear when the pumping mode responds in phase with the forcing, i.e. when the drop gets the most flattened shape as the vibrating bench reaches its highest - and most leftward - position. 

\begin{figure}
\includegraphics[scale=0.5]{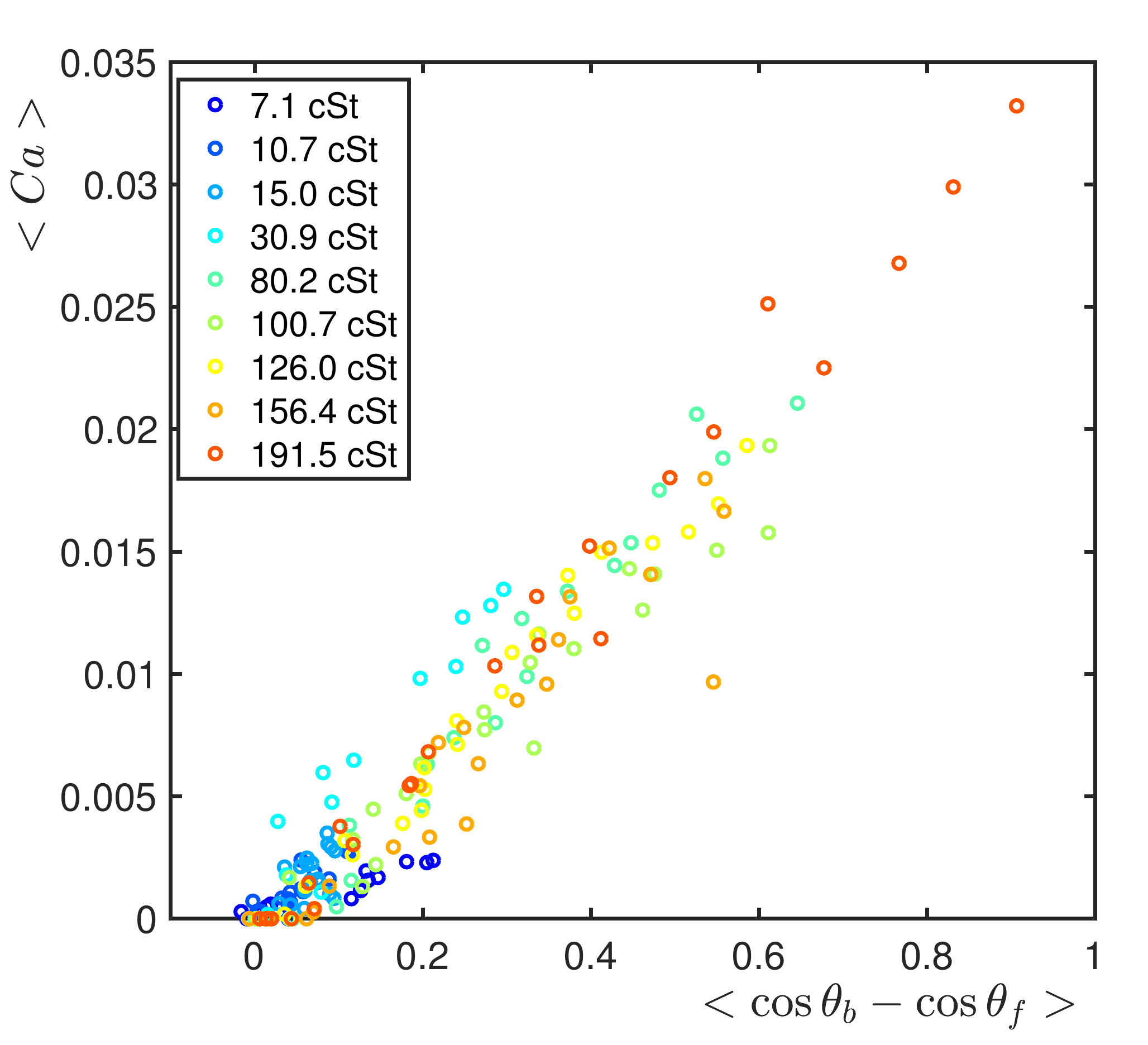}
\caption{Averaged capillary number versus the averaged difference of the cosines of back and front CAs. Experimental conditions are those of figure \ref{fig:influ_visco}.}
\label{fig:Ca_vs_cosCA_visco}
\end{figure}

Let us now come back to the second question stated at the end of the introduction, namely the relationship between the averaged drop dynamics and the unsteady one over one period. Indeed in previous experiments \cite{Brunet07}, it was possible to relate $<v>$ and values taken by $\theta_f$ and $\theta_b$ over one period. This relationship was rationalised by considering the averaged unbalanced Young force per unit length of the contact line, over one period :

\begin{equation}
F_y = \frac{\gamma}{T} \int_{0}^{T} (\cos \theta_b - \cos \theta_f) dt = \gamma < \cos \theta_b - \cos \theta_f >
\label{eq:young}
\end{equation}

\noindent and equating it with the friction force based on the averaged velocity $F_v = \eta <v>$. While this equilibrium neglects the fact that the velocity of the drop centre of mass does not advance with a constant velocity, stating $F_v = F_y$ turns out to be fairly correct in practice, although with a prefactor presumably of geometrical origin. In figure \ref{fig:Ca_vs_cosCA_visco}, we plotted the capillary number built on the averaged velocity $<Ca> = \frac{\eta <v>}{\gamma}$, versus $< \cos \theta_b - \cos \theta_f >$. We extracted the values of $\theta_b$ and $\theta_f$ corresponding with most of the measurements plotted in Fig.~\ref{fig:influ_visco}. The collapse of data for different values of $\eta$ is fair. It makes us confident that the relationship between $<v>$ and ($\theta_b$, $\theta_f$), despite the complexity of the time-dependence of the CAs, can be extended over a large range of viscosity and amplitudes.

\section{CONCLUSIONS}

We conducted a quantitative experimental study on the direction motion of sessile drops induced by slanted substrate vibrations, under frequencies typically lying in the range of the first inertio-capillary eigenmodes of the drop. In the aim to obtain quantitative trends, especially suitable to be compared with existing theories, our experiments spanned a large range of frequency and viscosity. To some extent, we also investigated the influence of the angle $\alpha$ between the normal to the substrate and the axis of vibration.

From our results, the main take-home messages are the following :

- for most experimental conditions, the averaged velocity $<v> \sim (A-A_{th})^{\chi}$, with $\chi = 1$. We then defined a natural mobility coefficient $s = \frac{\Delta <v>}{\Delta A}$ and the threshold for motion $A_{th}$.

- the influence of $\alpha$ is very sharp only in the ranges of roughly 10 degrees from the limits 0$^{\circ}$ and 90$^{\circ}$, and otherwise is rather marginal.

- the dependence of the mobility $s$ with $f$ suggests a complex interplay between rocking and pumping modes, in particular with an optimal mobility found when $f$ is slightly above $f_p$. 

- we observed a backward motion when $f$ is smaller than 0.25 $f_p$ or within a narrow domain around the higher order asymmetric mode $f_4$ (generally around or above 1.5 $f_p$). In the latter case, the drop's response exhibits a period halving, which originates from a parametric instability. 

- the dependence in viscosity shows that $s$ is almost independent on $\eta$ if $\frac{\delta}{h} < \delta^*_c \simeq 0.83$ and that $<v>$ scales as the inverse of $\eta$ if $\frac{\delta}{h} > \delta^*_c$, hence within a domain of higher viscosity.

- no clear trend could emerge from measurements of instantaneous contact-line velocities and dynamical CAs in terms of phase-shift of pumping and rocking modes with respect to the forcing amplitude. At a qualitative level, the results confirm that, at least when $\delta^* <$ 1 and for moderate to large $A$, the dynamical CAs take non-singled values with the contact-line velocity. Still, we could extract the dynamics of the pumping mode via the time-evolution of the basal diameter $(x_f - x_b)$: it exhibits monotonic increase of its peak-to-peak variations with $A$, while the time-lag $\Delta t /T$ shows weak dependence on $A$. This time-lag roughly increases with $f/f_p$, and reach a value close to one - i.e. the pumping mode in phase with the forcing vibration, near the optimum of mobility.

- the dimensionless velocity, the capillary number $<Ca>$, shows a linear relationship with the  capillary force averaged over one period, built on the averaged difference between the cosines of the back and forth dynamical angles $\theta_b$ and $\theta_f$. This confirms the trend from previous results \cite{Brunet07}, and consolidates them over a large range of viscosity. This scaling allows one to make the collapse of data for all the different tested liquid viscosities. 


\end{document}